\documentclass[]{aa}

\usepackage{graphicx}
\usepackage{txfonts}

\usepackage{natbib}  
\bibpunct{(}{)}{;}{a}{}{,} 

\newcommand{\Msun}{\ensuremath{\mathrm{M_{\sun}}}}
\newcommand{\mas}{\ensuremath{\mathrm{\, mas}}}

\newcommand{\kms}{km~s$^{-1}$}

\begin{document}
\title{Resolved astrometric orbits of ten O-type binaries.\thanks{Based on observations collected with the PIONIER/VLTI, AMBER/VLTI and GRAVITY/VLTI instruments at the European Southern Observatory, Paranal, under programs 087.C-0458, 087.D-0150, 087.D-0264, 090.D-0036, 090.D-0291, 090.D-0600, 091.D-0087, 091.D-0334, 092.C-0243, 092.C-0542, 092.D-0015, 092.D-0366, 092.D-0590, 092.D-0647, 093.C-0503, 093.D-0039, 093.D-0040, 093.D-0673, 094.C-0397, 094.C-0884, 189.C-0644, 60.A-9168, 096.D-0114.}\ \thanks{Table B.1 is available in electronic form at the CDS via anonymous ftp to cdsarc.u-strasbg.fr (130.79.128.5) or via\newline http://cdsweb.u-strasbg.fr/cgi-bin/qcat?J/A+A/}}

\author{
   J.-B.~Le~Bouquin\inst{1}
  \and H.~Sana\inst{2}
  \and E.~Gosset\inst{3}\fnmsep\thanks{F.R.S.-FNRS Senior Research Associate}
  \and M.~De~Becker\inst{3}
  \and G.~Duvert\inst{1}
  \and O.~Absil\inst{3}\fnmsep\thanks{F.R.S.-FNRS Research Associate}
  \and F.~Anthonioz\inst{1}
  \and J.-P.~Berger\inst{4}
  \and S.~Ertel\inst{10}
  \and R.~Grellmann\inst{7}
  \and S.~Guieu\inst{1,6}
  \and P.~Kervella\inst{5}
  \and M.~Rabus\inst{8,10}
  \and M.~Willson\inst{9}
}

\institute{
  Univ. Grenoble Alpes, CNRS, IPAG, F-38000 Grenoble, France
  \and 
  Institute of Astrophysics, KU Leuven, Celestijnlaan 200D, 3001 Leuven, Belgium
  \and
  Space sciences, Technologies and Astrophysics Research (STAR) Institute, Universit\'e de Li\`ege, 19C All\'ee du Six Ao\^ut, B-4000 Li\`ege, Belgium
  \and
  European Southern Observatory, Schwarzschild-Str. 2, D-85748 Garching bei M\"unchen, Germany
  \and
  Unidad Mixta Internacional Franco-Chilena de Astronom\'{i}a (UMI 3386), CNRS/INSU, France ; Departamento de Astronom\'{i}a, Universidad de Chile, Camino El Observatorio 1515, Las Condes, Santiago, Chile ; LESIA, UMR 8109, Observatoire de Paris, CNRS, UPMC, Univ. Paris-Diderot, PSL, 5 place Jules Janssen, 92195 
  \and
  European Southern Observatory, Alonso de Cordova 3107, Vitacura, 19001 Casilla, Santiago 19, Chile
  \and
  I. physikalisches Institut, Universit\"at zu K\"oln, Z\"ulpicherstr. 77, 50937 K\"oln, Germany
  \and
  Instituto de Astrofisica, Facultad de Fisica, Pontificia Universidad Cat\'olica de Chile, Chile
  \and 
  University of Exeter, Astrophysics Group, School of Physics, Stocker Road, Exeter EX4 4QL, UK
  \and 
  Max Planck Institute for Astronomy, K{\"o}nigstuhl 17, 69117 -
  Heidelberg, Germany
}

\offprints{J.B.~Le~Bouquin\\
  \email{jean-baptiste.lebouquin@univ-grenoble-alpes.fr}}
\date{Received 07/07/2016; Accepted 23/07/2016}
\abstract
{}
{Our long-term aim is to derive model-independent stellar masses and distances for long period massive binaries by combining apparent astrometric orbit with double-lined radial velocity amplitudes (SB2).}
{We followed-up ten O+O binaries with AMBER, PIONIER and GRAVITY at the VLTI. Here, we report on 130 astrometric observations over the last seven years. We combined this dataset with distance estimates to compute the total mass of the systems. We also computed preliminary individual component masses for the five systems with available SB2 radial velocities.}
{Nine of the ten binaries have their three-dimensional orbit well constrained. Four of them are known to be colliding wind, non-thermal radio emitters, and thus constitute valuable targets for future high angular resolution radio imaging. Two binaries break the correlation between period and eccentricity tentatively observed in previous studies. This suggests either that massive star formation produces a wide range of systems, or that several binary formation mechanisms are at play. Finally, we found that the use of existing SB2 radial velocity amplitudes can lead to unrealistic masses and distances.}
{If not understood, the biases in radial velocity amplitudes will represent an intrinsic limitation for estimating dynamical masses from SB2+interferometry or SB2+Gaia. Nevertheless, our results can be combined with future Gaia astrometry to measure the dynamical masses and distances of the individual components with an accuracy of 5 to 15\%, completely independently of the radial velocities.}

\maketitle

\section{Introduction}
\label{sec:intro}

Massive stars are key components of galaxies and, despite their importance in modern astrophysics, they remain incompletely understood. Direct measurements of some critical parameters remain scarce \citep{2003IAUS..212...91G,2012ASPC..465..275G}. In particular, masse estimates with an accuracy better than 5\% are still needed to challenge stellar evolution models \citep{2014A&A...570A..66S}.

So far, double-lined spectroscopic eclipsing binaries (SB2E) offer the most reliable constraints on the absolute masses. In the best cases, the formal precision on the masses can be as low as 1\%. Unfortunately, because of the decreasing alignment probability with increasing separation, the few known SB2E systems only give access to tight, short period binaries ($P < 10$\,d, typically). The complex influence of tidal forces on the internal structure is not negligible. Additionally, it is sometimes difficult to strictly exclude a prior mass-transfer event that could modify the amount of mass in each individual star. This is especially true because the evolutionary status of massive stars is not always easy to determine. These phenomena add significant degrees of uncertainty to the difficult task of calibrating evolutionary tracks from short period binaries \citep{2014A&A...570A..66S,2015ApJ...812..102A}.

In this context, being able to measure the true masses of longer-period, non-eclipsing systems could bring two fundamental advantages: (1) they alleviate the limitations imposed by the rareness of eclipsing systems, and (2) they avoid the uncertainties related to the effects of tides and  mass-transfer given the significantly larger physical separations. Long-period binary systems can be characterised through radial-velocity up to periods of about ten years \citep{2011IAUS..272..474S}. However the inclination of their orbital plane with respect to the line of sight needs to be obtained by other means. Optical long baseline interferometry (OLBI) has long been recognised as a promising solution \citep{boyajian:2007,kraus:2009apr}, although hampered by sensitivity and efficiency issues.

In recent years, the OLBI techniques underwent major improvements in performances, thanks to infrastructure and instrumental evolutions. From a handful of O-type stars (e.g. $\gamma^2\,$Vel, $\theta^1$~Ori~C, $\eta$~Car), the accessible sample increased to more than one hundred \citep{2014ApJS..215...15S}. Many massive binaries with periods in the range 100 to 5000\,days at 1 to 2\,kpc (a typical distance for nearby O stars) are now accessible by both radial-velocity studies and by the Very Large Telescope Interferometer (VLTI). A quantitative illustration of the overlap and complementarity between the techniques can be found in Figure~1 from \citet{2010RMxAC..38...27S}. This new situation has triggered a flurry of results in the field of massive stars \citep{2009A&A...506L..49M,Sana:2011,De-Becker:2012,2013A&A...553A.131S,2013A&A...554L...4S,2014A&A...572L...1S}.

In this context, we initiated a follow-up of long-period massive binaries that could be spatially resolved by optical interferometry. Most of our targets were selected from the catalogue of \citet{Sana:2012} with proven or suspected spectroscopic double-lined signature (SB2). By combining the SB2 radial velocity amplitudes with the size and inclination of the relative orbital motion on the sky, it is possible to determine the individual masses of each component and the distances to the systems (though see difficulties below). Our final aim is to confront evolutionary models and existing theoretical calibration laws such as that in \citet{2005A&A...436.1049M}.
Additionally, a complete knowledge of the orbital elements is of particular interest for specific objects, such as non-thermal radio emitters and hierarchical triples. In the long run, obtaining a good sample of the latter systems may indeed help understanding why over one third of massive systems are actually formed by three or more stars \citep{2014ApJS..215...15S}, and to constrain star formation theories.

This paper presents the observational results of the interferometric follow-up, and discusses the consequences with respect to the final goal of calibrating models. The second section details the sample, observations and analyses. The third section presents the results for each individual object. The fourth section discusses the potential of these results to provide reliable masses, with a special focus on perspectives offered by Gaia. The paper ends with brief conclusions.

\section{Observations and analyses}
\label{sec:observation}

\subsection{The sample}
The sample of stars is presented in Table~\ref{tab:sample}. We focus on long-period massive binaries with proved or suspected SB2 signature. The sample is restricted to southern targets with magnitude $H<8$ to match the location and sensitivity limit of the PIONIER instrument with the 1.8\,m Auxiliary Telescopes. We add three non-thermal radio emitters that were recently resolved as long-period binaries.

The initial sample also included HD152234, an O9.7I with a known spectroscopic period of $\approx125$ days \citep{Sana:2012}. However the spectroscopic companion was only marginally resolved in our first interferometric observations \citep{2014ApJS..215...15S}. The obtained accuracy with the current infrastructure would not allow us to reach our science goal for this system, and so we stopped interferometric monitoring of this object.

The individual masses listed in Table~\ref{tab:sample} are computed from the spectral types using the calibration from \citet{2005A&A...436.1049M}. Caution should be applied when using these masses, since their uncertainty is as high as 35 to 50\%. The reference for the spectral types (and/or directly the masses) are given in Column~4. The supposed association and corresponding distances are reported in Table~\ref{tab:sample}, with references in Column~7.

\begin{table*}\centering
\caption{Sample of targets.}  \label{tab:sample}
 \begin{tabular*}{0.92\textwidth}{ccccccccccc}
 \hline\hline\noalign{\smallskip}
   Target & Spectral & Masses & Refs. & Cluster & distance & Ref. & $H$ & $V$ \\
   name & type(s) & ($M_\odot$) & & & (kpc) & & (mag) & (mag)\\
 \noalign{\smallskip}\hline\noalign{\smallskip}
\object{HD 54662}    &  O6.5~V + O9~V       & 29+18 & (a) & CMa OB1 & $1.1\pm0.1$ & (1) & 6.172 &  6.212\\
\object{HD 93250}    & O4~III + ?           & 49+? & (b) & Carina Ass. & $2.35$ & (2) & 6.720 &  7.365\\
\object{HD 150136}   & (O3~V + O6~V) + O7~V & (81)+27 & (c) & NGC 6193 & $1.32\pm0.12$ & (3) & 5.090 & 5.540\\
\object{HD 152233}   & O5.5~III + O7.5~III/V& 38+30 & (d) & NGC 6231 & $1.52$ & (4,5) & 6.145 & 6.556\\
\object{HD 152247}   & O9~III + O9.7~V      & 23+16 & (d) & NGC 6231 & $1.52$& (4,5) & 6.614 & 7.172\\
\object{HD 152314}   & O8.5~V + B2~V & 19+10 & (d) & NGC 6231 & $1.52$& (4,5) & 7.243 & 7.866\\
\object{HD 164794}   & O3.5~V + O5.5~V   & 50+34 & (e) & NGC 6530 & $1.25\pm0.1$ & (6) & 5.748 & 5.965\\
\object{HD 167971}   & (O7.5III + O9.5III) + O9.5~I & (50)+26 & (f) & NGC 6604 & $1.75\pm0.2$ & (7) & 5.315 & 7.479\\
\object{HD 168137}   & O7~V + O8~V & 22+24 & (d) & NGC 6611 & $1.8\pm0.1$ & (8) & 7.683 & 8.945 \\
\object{CPD-47 2963} & O5~I + ? & 51+? & (g) & Vel OB1 & $1.3$ & (9) & 6.060 & 8.45 \\
 \noalign{\smallskip}\hline
 \end{tabular*}
\tablebib{   (a) \citet{boyajian:2007apr} ; (b) \citet{Maiz-Apellaniz:2004} ; (c) \citet{2013A&A...553A.131S} ; (d) \citet{Sana:2012} ; (e) \citet{Rauw:2012} ; (f)~\citet{De-Becker:2012} ; (g) \citet{2014ApJS..211...10S} ; 
(1)~\citet{2000MNRAS.312..753K} ; (2)~\citet{2006ApJ...644.1151S} ; (3) \citet{1977A&AS...30..279H} ; (4) \citet{2005A&A...441..213S} ; (5)~\cite{2013AJ....145...37S} ; (6) \cite{2005A&A...430..941P} ; (7) \citet{2013MNRAS.436..750I} ; (8) \citet{2006A&A...457..265D} ; (9) \citet{2006PASA...23...50B} }
\end{table*}

\subsection{Observations and data reduction}
Most interferometric data were obtained with the PIONIER\footnote{http://ipag.osug.fr/pionier} combiner \citep[][]{Le-Bouquin:2011} and the four 1.8\,m Auxiliary Telescopes of the VLTI \citep{Haguenauer:2010}. To calibrate the fringe visibilities and closure phases, we interleaved observations of KIII stars between those of the O-type stars. These reference stars were found with the \texttt{SearchCal}\footnote{http://www.jmmc.fr/searchcal} software \citep{Bonneau:2011a}. These K giants have apparent diameters smaller than $0.5\,\mas$ therefore providing calibration with 5\% accuracy even on the longest baselines of VLTI. Data were reduced and calibrated with the \texttt{pndrs}\footnote{http://www.jmmc.fr/pndrs} package \citep[][]{Le-Bouquin:2011}. Each observing block (OB) provides five consecutive files, each containing six visibilities and four closure phases dispersed over three spectral channels across the H band. The dataset is complemented by about ten observations performed with the AMBER combiner and three 8\,m Unit Telescopes, mostly before 2011. These data were reduced and calibrated with the \texttt{amdlib-3} package \citep{tatulli:2007mar,chelli:2009aug}. Finally, two 2016 observations were obtained during the science verification of the GRAVITY instrument \citep{2011Msngr.143...16E}. They were reduced and calibrated by the standard pipeline \citep{2014SPIE.9146E..2DL}. The description of the GRAVITY instrument and its first results will be presented in detail in a forthcoming paper from the GRAVITY Collaboration (corresponding author: F. Eisenhauer, e-mail: eisenhau@mpe.mpg.de).

\subsection{Resolved astrometry fitting}
 The interferometric observations were adjusted with a simple binary model composed of two unresolved stars. This is a valid assumption because the expected diameters of the individual components ($<0.2$\,mas) are unresolved, even with the longest VLTI baselines. The free parameters are the two coordinates of the apparent separation vector (East and North, in milliarcsecond) and the flux ratio $f_H$ between the secondary and the primary. The latter is considered constant over the H band and over the different epochs. Given the long orbital periods of these binaries ($P \approx 0.5 - 15$~yr), we neglected the orbital motion in the course of a single night.

We here report $\approx$$130$ observations, a tenth which have already been presented in \citet{2014ApJS..215...15S}. The journal of the observations and the astrometric measurements can be found in Appendix~\ref{tab:obslog}. It includes the best-fit binary position and its associated uncertainty ellipse for each modified julian date (MJD). The total execution time for a single resolved astrometric observation is about 30\,min, including calibrations.

%
%
%

\subsection{Orbit determination}
\label{sec:orbit}
We use a Levenberg-Marquardt method to find the best orbital solutions. A more complex search is not necessary because the temporal sampling is dense enough for the period to be determined unambiguously. We adjust simultaneously the resolved astrometric positions and the radial velocities, when available. This is similar to what is presented in \citet{Le-Bouquin:2013}. There are ten independent parameters to fit, namely the orbital period $P$, the time of periastron passage $T$, the size of the apparent orbit $a$, the orbital eccentricity $e$, the inclination $i$, the longitude of ascending node $\Omega$, the argument of periastron of the secondary $\omega$, the semi-amplitudes $K_a$ and $K_b$ of the primary and secondary radial velocity curves, and the systemic heliocentric offset velocity $g$.

The ascending node is the node where the motion is directed away from the observer, that is when the passing star has a positive radial velocity. The argument of periastron is the angle between the ascending node and the periastron of the secondary, measured in the plane of the true orbit and in the direction of the motion of the secondary. The longitude of ascending node, measured in the plane of the sky, is counted positively from North to East ($\Omega = 90\,$deg). An inclination larger than 90 deg indicates an astrometric orbit covered clockwise.

We suppose the primary component for radial velocity (most massive) corresponds to the astrometric primary (brightest in H band). When only SB1 radial velocities are available, we assume that they represent the radial velocities of the primary. When no radial velocities are available, we enforce $0 < \Omega < 180\,$deg (using $\Omega + 180\,$deg and $\omega + 180\,$deg gives the same fit to the astrometric orbit alone).


%
%
%
%

\subsection{Uncertainties}
\label{sec:uncertainties}

The uncertainties are propagated numerically. A series of adjustments is performed while the data are corrupted by a random additive noise corresponding to their uncertainties. The quoted uncertainty on the orbital parameter is the root mean square of these various realisations. This method also allows us to explore the possible correlation between the parameters. A systematic uncertainty of 1.5\% should be added on the size of the apparent orbit $a$. It comes from the calibration of the effective wavelength of the PIONIER instrument that was used to obtain almost all resolved astrometric points. This translates into a ``plate scale error'' through the interferometric equation.

\subsection{Physical quantities}
Once the orbital parameters are known, additional physical quantities can be derived. We define two strategies: one fully independent of the radial velocities and one fully independent of the adopted distance. In the first approach, we impose the distance $d$ from external constraints such as membership to a known cluster or association (see Table~\ref{tab:sample}). It allows to compute the physical size of the orbit $a_p = d\,a$. The total mass $M_t$ of the system is then derived with Kepler's third law:
\begin{equation}
M_t = \frac{ 4\pi^2\;a_p^3}{\mathrm{G}\,P^2}, \label{eq:k3}
\end{equation}
where $\mathrm{G}$ is the gravitational constant. In the second approach, we use the radial velocity amplitudes, when available, to determine the physical size of the orbit, 
\begin{equation}
      a_p = \frac{P\,(K_a+K_b)}{2\pi}\,\frac{\sqrt{1-e^2}}{\sin(i)}
\end{equation}
and thus the total mass from Eq.~\ref{eq:k3}. The distance of the system is $d=a_p/a$ and the individual masses are given by:
\begin{equation}
     M_a = \frac{K_b}{K_a+K_b}\,M_t \ \ \ \ \ \ \ \ \ \ \ \ \ ;
     \ \ \ \ \ \ \ \ \ \ \ \ \ M_b = \frac{K_a}{K_a+K_b}\,M_t\;\;\;.
\end{equation}



\section{Results}
\label{sec:results}

The results for each binary of the sample are displayed in Appendix~\ref{sec:individualresults}. Here we discuss the specificities of each system.

\subsection{HD54662}
This massive binary is a known runaway star producing a bow shock as its winds interact with the interstellar medium \citep{2012A&A...538A.108P}. It was first recognised as a promising target for mass determination by \citet{boyajian:2007}. The three-dimensional orbit is now fully constrained by the resolved astrometric observations. However the orbital period of almost six years is completely different from the preliminary estimation from radial velocities ($557\,$d). It illustrates that resolved astrometry is less prone to phase ambiguities. We have attributed the historical SB1 radial velocities to the primary as they closely match its SB2 amplitude. We did so to confirm that the historical spectroscopic binary corresponds to the interferometric binary.

The total mass is in agreement with expectations from spectral types. However the individual masses are still uncertain because of the lack of accurate SB2 radial velocity measurements. The preliminary estimates for $K_a$ and $K_b$ from \citet{boyajian:2007} also lead to unrealistic masses, and should then be considered with caution.

Finally, the eccentricity is rather small considering the period of six years.

\subsection{HD93250}
This target is a non-thermal radio emitter that was first spatially resolved as a binary by \citet{Sana:2011}. Here we present additional observations that fully constrain its astrometric orbit. The projected semi-major axis of 1.2~mas is at the limit of the spatial resolution of the VLTI and some measurements have large relative errors. Accordingly, it has the shortest period of our sample ($P = 194.3$~d). The orbital period is still long enough to avoid tidal interaction and circularisation. The total mass of the system points toward a pair of similar high mass stars (O4+O4), as do the measured flux ratio near unity. The system is also known for the presence of non-thermal radio emission (colliding winds).

There is no radial velocity variation reported for this star \citep{2009MNRAS.398.1582R}, but given its short period it was worth obtaining a full orbit to guess the expected signal. However, because of the small inclination, the expected radial velocity variations due to the orbital motions are about 20~\kms. It is probably impossible to disentangle this from the intrinsic line width, as first suggested in \citet{Sana:2011}. It illustrates well the detection biases affecting binaries with (quasi) identical components.

While HD~93250 is inadequate for direct mass determination from resolved astrometry and radial velocities, constraints on the orbital parameters are valuable to properly interpret the non-thermal radio emission. A significant eccentricity was also required to explain the variations observed in the X-ray flux and is confirmed by this study ($e=0.22\pm0.01$).

\subsection{HD150136}
This is a known hierarchical triple system and one of the most massive stellar systems known \citep{Mahy:2012}. The first spatially resolved observations of the outer pair were discussed by \citet{2013A&A...553A.131S} and \citet{2013A&A...554L...4S}. Here we present additional observations and an improved orbit. The new orbit has a slightly longer period (3065~d vs.\ 3008~d), slightly lower eccentricity (0.68 vs.\ 0.73) and slightly different absolute masses. All stellar and orbital parameters are within $1\sigma$ of the previously published values. The accuracies are, however, improved. In Table~\ref{fig:HD150136} (right), component $a$ corresponds to the inner 2.7~d binary while component $b$ is the O6.5-7~V outer companion.

The three-dimensional orbit of the 8.4 year outer system is now uniquely constrained thanks to our long term follow-up. Still, it would benefit from few additional spatially resolved observations in the coming years to further reduce the uncertainties on the absolute masses.

Regarding the inner, short period binary: accounting for the minimal masses from \citet{Mahy:2012} and our new total mass of 88~\Msun, we obtain an inclination of $i=53$\degr\ and thus individual masses of 54 and 34~\Msun. Again, these values are at the lower bound of the $1\sigma$ confidence interval from the preliminary study of \citet{2013A&A...553A.131S}.

The distance estimate from the combined fit favours the new distance to the NGC\,6193 cluster ($1.15\,$kpc) from \citet{2005A&A...438.1163K} compared to the older one ($1.32\,$kpc) from \citet{1977A&AS...30..279H}.

\subsection{HD152233}

The well constrained three-dimensional orbit reveals an almost edge-on eccentric binary. The measured properties correspond with those of the long period SB2 binary discussed by \citet{2008MNRAS.386..447S}. The total mass of the system ($\approx 45\,M_\odot$) is significantly lower than the expectation from the spectral types ($38+30\,M_\odot$, Table~\ref{tab:sample}). Together with the small flux ratio, it points towards a lower mass secondary, most likely of a later spectral type as we discuss further below.

The individual masses ($\approx 2$~\Msun) and distance (700~pc) derived from the combined astrometric and RV fit are unrealistic. They cannot be reconciled with massive stars, even considering large uncertainties. A visual inspection of the UVES spectra shows that the spectral lines are never fully disentangled by the radial velocity shifts. The radial velocities of the secondary may be systematically underestimated in the spectral analysis, thus lowering the total mass and the distance of the system. If we impose a distance of $\approx1.5$\,kpc (thus somehow imposing the total mass to be $\approx 45\,M_\odot$), the fit of the radial velocities of the secondary yields a systematic residual of about 15\,\kms, that is of similar amplitude as the uncertainties. The individual masses rise to $31$ and $12\,M_\odot$. These are more realistic values, also compatible with the small flux ratio measured. We conclude that the radial velocities are probably faulty.

Given the high inclination, the closest apparent approach is $0.24\pm0.05\,$mas. However there is little chance that this system could be a long-period eclipsing binary because the expected apparent stellar diameters are about $0.17\,$mas.

\subsection{HD152247}
As seen in the pair diagram, the three-dimensional orbit is well constrained. It corresponds to the long period, highly eccentric spectroscopic binary discussed in \citet{Sana:2012}.

The total mass is compatible with the masses expected from spectral types. The measured flux ratio agrees with the expectation for late O-type stars with luminosity classes III and V. A small amount of accurate SB2 data should allow to obtain individual mass estimates independently from the distance. Attributing the reported SB1 radial velocities to the primary, we expect SB2 velocity semi-amplitudes of $\approx 61\,$\kms\ for the secondary (fainter and less massive). Thus HD152247 is confirmed as a good candidate for precise mass determination.

\subsection{HD152314}
The three-dimensional orbit is well constrained. It corresponds to the long period spectroscopic binary discussed in \citet{Sana:2012}. 

The total mass and the measured flux ratio are roughly compatible with the masses expected from spectral types. Attributing the reported SB1 radial velocities to the primary, we expect SB2 velocity semi-amplitudes of $\approx 30\,$\kms\ for the B-type secondary.

\subsection{HD164794\ \ \  (9~Sgr)}
The three-dimensional orbit is well constrained. Our combined fit confirms that the resolved pair corresponds to the long-period SB2 presented in \citet{Rauw:2012,2016arXiv160306745R}. We had to introduce a constant RV shift of $10.0$\,\kms\ between the two components in order to achieve a correct fit of the SB2 radial velocities. This is not unusual for very massive stars and is thought to be an effect of the stellar winds. The combined fit is convincing but this result is the most puzzling of our sample.

First, placing the system at a distance of $d \approx 1.25$\,kpc makes the total mass compatible with the expected spectroscopic one ($\approx 80\,M_\odot$). Thus, in the long controversy about the distance to NGC6530, and assuming the total mass from spectroscopy is correct, our observations confirm the distance from \citet{2005A&A...430..941P} and \citet{2005A&A...438.1163K}. This is significantly lower than the $1.78\pm0.08\,$kpc assumed by \citet{2016arXiv160306745R} and based on \citet{2000AJ....120..333S}. 

Although it makes no assumption on the distance, the combined fit between radial velocities and the resolved astrometry does not solve the issue. The deduced individual masses and the distance are even smaller. We note that, fitting the same RV data, \citet{2016arXiv160306745R} obtain similar radial velocity amplitudes. They conclude that the masses were compatible with expectation, but  assuming an inclination of $45\pm1$\,deg. Looking at the resolved astrometry in Fig.~\ref{fig:HD164794}, it seems hard to significantly lower the inclination of our best-fit of $86$\,deg.
In fact, such a low inclination is definitely ruled out by the VLTI observations. Thus we are left with a clear discrepancy between the masses expected from spectral types, the best-fit semi-amplitudes of the radial velocity curves and the distance to the system.

To further explore, we ran a combined fit of astrometry and radial velocity, but imposing the distance to be $1.25$\,kpc (thus somehow imposing the total mass to be $\approx 75\,M_\odot$). The agreement to the radial velocities is obviously significantly degraded, but the systematic residuals are never larger than $20\,$\kms. Altogether, we believe this is our best proxy for this system ($M_a\approx 40\,M_\odot$ ; $M_b\approx 33\,M_\odot$ ; $d\approx 1.25\,$kpc ; $i\approx 86\,$deg). A careful re-analysis of the radial velocity amplitudes taking into account the possible systematic biases is required to understand the disagreement among the observations.

Finally, even if the orbit is almost edge-on, the system is unlikely to be eclipsing. The closest apparent approach is $0.39\pm0.1\,$mas, for estimated stellar diameters of 0.14\,mas.

\subsection{HD167971}
This is a known hierarchical triple system in which we observe the orbital motion of the third component around the short period eclipsing pair. This outer pair is also the longest-period system in our sample. It was first resolved by \citet{De-Becker:2012}. Here we provide additional interferometric observations.

Any period between 15 and 25 years could easily match the astrometric observations alone. Our temporal coverage is still not sufficient to obtain a definite estimate of the period, and thus of any of the orbital elements. This should be easily tackled with additional interferometric points in the coming years.

Interestingly, the close eclipsing pair is intensively studied and modelled by \citet{2013MNRAS.436..750I}. By measuring the times of minimum light, they infer a period of $P=21.7\pm0.7\,$years for the outer pair. This value is compatible with the long term radio variability, interpreted as the signature of a colliding wind region between the winds of the inner binary and that of the third component \citep{2007A&A...464..701B}. We thus add these constraints ($P=21.7\pm0.7\,$years, $e=0.53\pm0.05$) in our fit of the astrometric orbit. The total mass of the system inferred from the orbit is in good agreement with the values given in Table~\ref{tab:sample}, even though one has to notice the large uncertainty on this quantity quoted in Fig.~\ref{fig:HD167971}.

The inclination of the outer orbit ($180-i = 35\,$deg) is very different than that of the eclipsing pair ($73\,$deg).

We found an average flux ratio of $0.78$, similar to the one first reported in \citet{De-Becker:2012}. Our interferometric measurements alone could not lift the ambiguity in the association of the resolved components to the close eclipsing binary and to the distant third star.

\subsection{HD168137}
The three-dimensional apparent orbit corresponds to the long period binary discussed from radial velocities by \citet{2009MNRAS.400.1479S}. The preliminary orbit has an impressive eccentricity of $0.90$. We checked various ways to explain this eccentricity with biases in the observations but could not find any plausible explanations. Additional observations in the coming years will firmly establish this peculiar eccentricity, and lift any remaining degeneracies. The next periastron passage is expected in July 2020.

The total mass is unconstrained because of the still poor precision on the size of the orbit ($\pm 50$\%). Again, a few additional interferometric observations will easily tackle this issue. Moreover the few existing SB2 radial velocities are very difficult to model, except for the well separated spectra. The other measurements might be a mixture of the two components, thus cancelling each other out. This is especially possible because the two stars have similar brightness ($f_H=0.77$) and similar line depths and widths \citep[see Fig.~5 in ][]{2009MNRAS.400.1479S}. We have ignored these radial velocities for the fit. The radial velocity elements are consequently degenerated, and the individual masses are unconstrained.

The most interesting result of this system is its huge eccentricity that could potentially make it a long-period eclipsing system. The expected apparent closest approach is $0.15\pm0.13$\,mas. This value thus has some probability of being lower than $0.1\,$mas, while the stellar diameters should be of this size. Altogether this system deserves additional monitoring to better determine its inclination, eccentricity and time of closest approach (so far known within $\pm200$ days).


\subsection{CPD-47 2963}
This non-thermal radio emitter was first spatially resolved as a binary in the course of the SMASH+ survey \citep{2014ApJS..215...15S}. Here we present additional observations and the first orbit. The three-dimensional orbit is constrained but would benefit from few additional interferometric observations to lift the remaining degeneracy between $\Omega{}$ and $\omega{}$ and between $a$ and $i$. The total mass is at the lower bound of the expectation from the spectral type, although still uncertain.

CPD-47\ 2963 has been reported as SB1 with a preliminary period of 59 days and a semi-amplitude of 9\,\kms\ \citep[OWN survey pre-publication,][]{2014ApJS..211...10S}. Considering the eccentricity and the period of the interferometric binary ($e=0.66$, $P=655$\,d), a stable hierarchical system seems unlikely according to the criterion quoted by \citet{2004RMxAC..21....7T}:
\begin{equation}
P_{\mathrm{out}}\,(1 - e_{\mathrm{out}})^3 > 5\,P_{\mathrm{in}}\ \ \ \ \ .
\end{equation}
 It is more plausible that all observations refer to the same pair. 

\section{Discussion}
\label{sec:discussion}

\subsection{Orbit of non-thermal radio emitters}
Non-thermal (synchrotron) radio emitters belong to the category of the so-called particle accelerating colliding-wind binaries (PACWBs), a first catalogue of which is published by \citet{2013A&A...558A..28D}. These objects are of great interest for understanding non-thermal physics. Particle acceleration arises from the strong hydrodynamic shocks in the colliding-wind region, providing an opportunity to investigate particle acceleration in environments different from, for example, supernova remnants.

PACWBs are intrinsically variable sources on a timescale corresponding to the orbital period, with an observed variable behaviour depending on the orientation of the orbit. The determination of their orbital elements is thus especially important.

Short period (i.e. a few days) systems are not suitable for the investigation of non-thermal processes. First of all, the acceleration of electrons in short period binaries is expected to be severely inhibited due to the highly efficient cooling by inverse Compton scattering of photospheric photons \citep{2007A&ARv..14..171D}. Second, the large dimensions of the radio-sphere \citep[up to several hundreds of R$_\odot$, see e.g. ][]{Mahy:2012} also prevent any potential synchrotron radio emission produced in the colliding-wind region to escape if the two stars are too close to each other. Long-period systems are thus better suited, but a proper knowledge of their orbit is often lacking. Those systems are challenging to follow-up with radial velocities due to large amounts of time needed and to the small amplitudes of the radial velocity variations.

Our sample includes one relatively short-period (HD93250) and three long-period non-thermal radio emitters. HD164794 and CPD-47\ 2963 have their three-dimensional orbits fully mapped. HD167971 will be completed soon. The updated number of PACWBs with at least partly determined orbits is now 14, compared to the 11 initially mentioned by \citet{2013A&A...558A..28D}. The full determination of their orbits is necessary to define adequate strategies for future high angular resolution radio observations using very long baseline interferometry (VLBI). It permits to select adequate orbital phases for observation, and allows for an adequate interpretation of radio images of the synchrotron emission region afterwards. Finally, potential future observations of PACWBs in gamma-rays will require a careful selection of the orbital phases as well. 



\subsection{Periods, inclinations and eccentricities}
Eight binaries of the sample fall within the ``interferometric gap'' defined by \citet{1998AJ....115..821M} where the periods are too long to be easily catched in radial velocity snapshot surveys but the apparent separations are too small to be resolved by imaging surveys. However most of them still have detectable radial velocity variations (except HD93250 which remained hidden because of its small inclination and nearly twin components).

Our target selection is biased toward large inclination because we favoured systems with detected SB2 lines (at least tentatively). Accordingly, the seven binaries with reported radial velocity variations have inclinations less than 15\,deg from edge-on. However, given the long periods involved, only one of them has some chance of being eclipsing.

\begin{figure}  \centering  \includegraphics[scale=1]{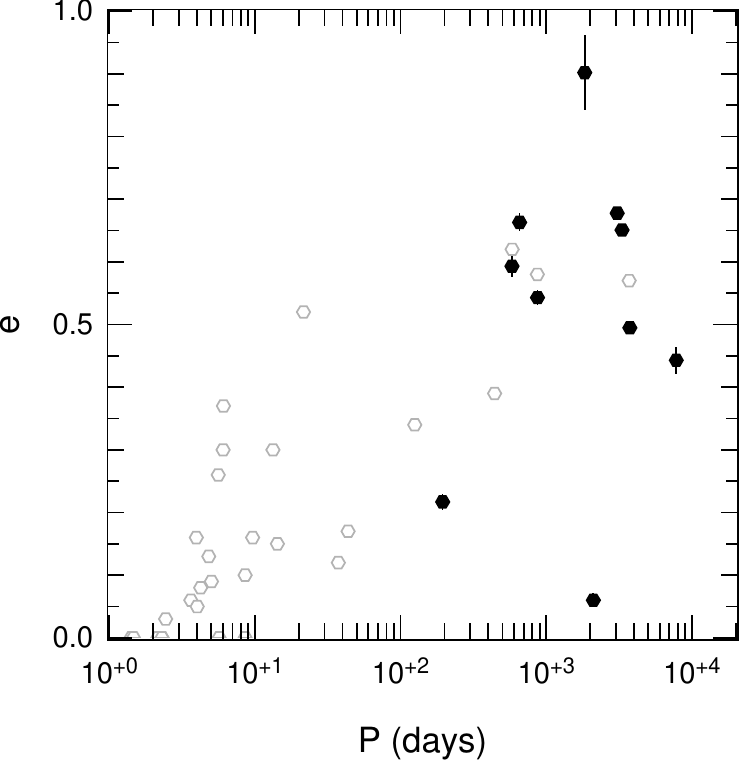}
 \caption{Eccentricity versus period of the best-fit solution. Filled black symbols are the binaries presented in this study. Open grey symbols are the SB2 listed by \citet{Sana:2012} with known eccentricity (included those of the present study).}
 \label{fig:eP} \end{figure}

Figure~\ref{fig:eP} shows the period-eccentricity diagram. Nine of the ten binaries in our sample have a well-determined eccentricity, and this parameter is partially constrained for the remaining one. It is known that the radial velocity technique requires a good sampling near the periastron passage in order to constrain orbits with high eccentricity. Resolved astrometry is more robust to high eccentricity because it measures the integrated orbital motion, and not its derivative.

It is noteworthy that the eccentricities are distributed over a large range. For example, there are two systems with periods near 2000\,days: one of them has the highest eccentricity of the sample ($0.90$) while the other has the lowest eccentricity ($0.06$). Interestingly, these two objects fall in parts of the diagram that were left empty before. They completely break the correlation between period and eccentricity tentatively observed in previous studies (see open symbols in Fig.~1). This finding suggests that the current lack of long-period systems with low ($e<0.25$) or high ($e>0.75$) eccentricities results from observational biases. In the former case, the small radial velocity variations are easily hidden by the typical rotational broadening of the stellar lines in massive stars \citep[see the discussion in][]{Sana:2011}. In the later case, the radial velocity variations potentially occur over a small fraction of the orbit. The difficulty of catching the periastron passage and the long timescales between two periastron passages make these systems challenging to detect and characterise.

The fact that such a variety of eccentricities co-exist indicates either that massive star formation produces a wide range of systems, or that several binary formation mechanisms are at play.

\subsection{Increasing the interferometric sample}
Recently, \citet{2014ApJS..215...15S} resolved about 20 additional systems within a similar range of separations. The expected periods should be similar, and a long-term interferometric follow-up has been initiated. Another interesting target is Herschel 36 (Aa-Ab), which has a period of 500\,d and was recently spatially resolved by VLTI as well \citep{2014A&A...572L...1S}.

It would be interesting to reach systems with slightly shorter periods, because they are more frequent and have larger radial velocity amplitudes. At VLTI, the limitation comes from the length of the interferometric baselines (150\,m, 1\,mas at H-band). The next generation GRAVITY/MATISSE instruments will not change the game, but simply participating in this follow-up initiated with PIONIER. The only way to improve for VLTI is to use shorter wavelengths.

The Center for High Angular Resolution Array (CHARA) has significantly longer baselines \citep[330\,m, 0.5\,mas at H-band,][]{ten-brummelaar:2008jul}. However, efficient follow-up of binaries are nowadays limited to stars brighter than 5\,mag. This should improve in the coming years with the arrival of new detectors and the installation of adaptive optics \citep{2016AAS...22742702T}.

\subsection{Preliminary masses}
\label{sec:preliminarymasses}
The distances to the systems are known with a precision of 5 to 15\,\%. For most binaries, it dominates the error budget of Eq.~\ref{eq:k3}. The total masses are thus known with a precision of 15 to 50\,\%. Only HD167971 and HD168137 have the precision on their total mass limited by the knowledge of the period and/or size of the orbit. For these two stars, this will be tackled in the coming years with some additional interferometric observations at VLTI.

The preliminary masses estimated from the SB2 radial velocities are puzzling. One system has realistic masses (HD150136), one has masses obviously overestimated (HD54662) and two have masses unrealistically low (HD152233 and HD164794). We speculate that the measured SB2 radial velocities are affected by systematic biases, phased with the orbital motion. The origin of such biases can be twofold:\begin{itemize}
\item An astrophysical effect makes the line-forming region appear differently depending on the orbital phase. A known example is the Struve–Sahade effect \citep{2007A&A...474..193L,2012A&A...537A.119P}. However this hypothesis is hard to support because of the very large physical separations of the binaries in our sample. Another effect could be related to the wind-wind collision in eccentric binaries. In HD~164794 for instance, \citet{Rauw:2012} described that various primary lines might behave differently around periastron passage. Depending on the relative weighting of the lines in the final orbital solution, this could bias the estimation of the radial velocity amplitude that critically depends on the measurements at periastron.

\item The spectral modelling is corrupted by the entanglement of the lines of each component, which varies according to the orbital phase. This is obviously the case when the true line profiles of each component are incorrectly modelled.
This second hypothesis is further supported by the fact that the only system for which the masses are found to be in agreement is also the only SB3. The orbital motion of the inner pair introduces large radial velocity shifts ($>200\,$\kms). This is well illustrated in Fig.~1 from \citet{Mahy:2012}. These enhanced shifts permit them to disentangle the line profile of the third component, and thus to model it adequately in most blended spectra.
\end{itemize}

\subsection{Perspectives for accurate SB2 velocities}
The limiting factor for mass determination of these long-period binaries is the accuracy of the SB2 radial velocity amplitudes (Sec.~\ref{sec:preliminarymasses}), and especially their sensitivity to the adopted line profiles. A possible mitigation is to directly fit the entire set of spectra by a combination of stellar line profiles, shifted accordingly to the known orbital parameters. The three remaining free parameters, namely $K_a$, $K_b$ and $g$, can be estimated without extracting radial velocities for each phase (note that $g$ may have to be 
be left free for each individual line). This approach makes optimal use of the information included in the blended spectra. Varying the stellar line profile parameters allows to investigate the final mass accuracies, including systematics.

To reduce this uncertainty, one should input the best possible individual line profiles. Future spectroscopic observations should focus on the predicted phases of maximal velocity, to provide the best possible disentangled spectra. Table~\ref{tab:rvmax} summarises the expected maximal velocity shifts. Five binaries have predicted shifts of the order of 100\,\kms\ or larger, thus enabling decent disentangling and model-independent masses. Disentangling will be much more difficult for the five other systems.

\begin{table}\centering
\caption{Velocity shifts $K_a+K_b$ predicted from the orbital parameters $a$, $P$, $i$ and $e$ of this study and the distances $d$ to the systems listed in  Table~\ref{tab:sample}. \label{tab:rvmax}}
 \begin{tabular*}{0.47\linewidth}{lc}
 \hline\hline\noalign{\smallskip}
   Target & $(K_a+K_b)$ \\
   name & (\kms) \\
 \noalign{\smallskip}\hline\noalign{\smallskip}
HD 54662 & 57 \\ 
HD 93250 & 61 \\ 
HD 150136 & 103 \\ 
HD 152233 & 94 \\ 
HD 152247 & 99 \\ 
HD 152314 & 46 \\ 
HD 164794 & 80 \\ 
HD 167971 & 28 \\ 
HD 168137 & 167 \\ 
CPD-47 2963 & 61 \\ 
 \noalign{\smallskip}\hline
 \end{tabular*}
\end{table}

\subsection{Perspectives with Gaia}
At the considered range of distances, Gaia will deliver exquisite parallax accuracy ($<2$\%), especially because the binary effect can be adequately subtracted thanks to the knowledge of the astrometric orbital parameters presented in this study. Combined with the size of the apparent orbits and the periods, we expect the total masses to be measured with an uncertainty of 5\%, following Eq.~\ref{eq:k3}. For most systems, the error budget will be equally shared between the interferometric resolved astrometry and the Gaia distance. These measurements of the total masses will be completely independent from radial velocities.

Gaia will also constrain the orbital photocentre displacement $\mu_G$. This quantity can be extracted even if the orbital period is poorly sampled by Gaia because of the knowledge of all orbital parameters. It permits calculating the mass ratio $q$:
\begin{equation}
  q = \frac{\mu_G f_G + \mu_G + a f_G}{a -\mu_G f_G - \mu_G} \label{eq:qFromGaia}
\end{equation}
where $a$ is the (known) size of the apparent orbit and $f_G$ the flux ratio in the Gaia G band. Interestingly, $f_G$ is well approximated by the (known) flux ratio in the H band $f_H$ because hot stars have their optical and near-infrared spectra in the Rayleigh-Jeans regime.

\begin{figure}  \centering  \includegraphics[scale=1]{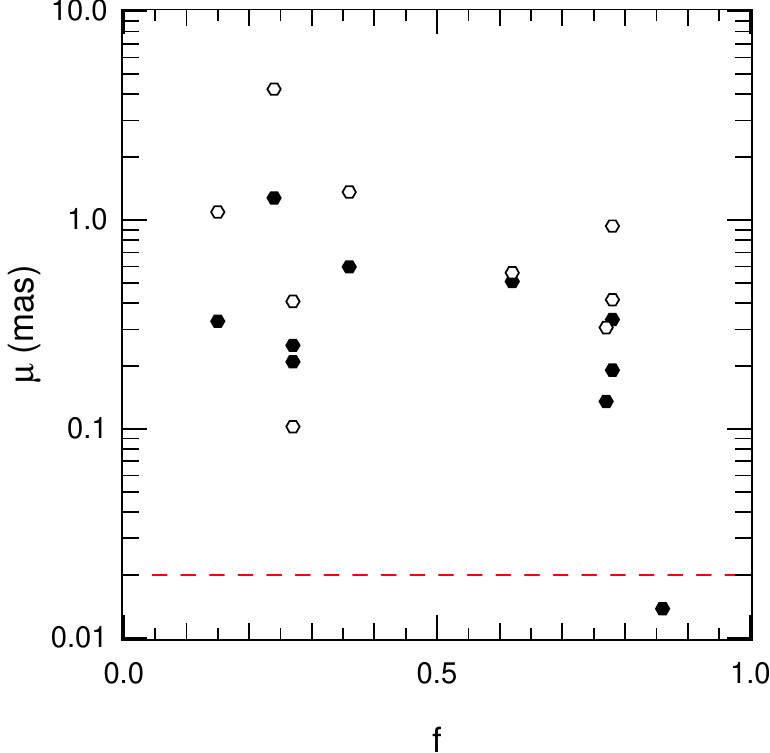}
 \caption{Predicted photocentre displacement versus the flux ratio for the ten binaries of the sample. The mass ratios were estimated either from the flux ratio (assuming main sequence stars, filled symbols), or derived from the individual spectral types when available (open symbols). The dashed red line is for the expected Gaia performances.}
 \label{fig:muf} \end{figure}

In order to test feasibility, we computed the expected photocentre displacement $\mu_G$ from the orbits derived in this study. This was done by inverting Eq.~\ref{eq:qFromGaia}, guessing the expected mass ratio. We used two approaches, both relying on the calibrations from \citet{2005A&A...436.1049M}. We first converted the observed flux ratio $f_H$ into mass ratio. For a pair of main-sequence stars, the calibration is well approximated by $q=f_H^{0.7}$. The exponent is even lowered for a massive, evolved primary with a less-massive main sequence secondary. In a second approach, we simply use the expected mass ratio from the individual spectral types when available (see Table~\ref{tab:sample}). The computed $\mu_G$ are reported in Fig.~\ref{fig:muf}. The predicted displacements range from 0.1\,mas to 1\,mas. This is a factor ten to a hundred larger than the Gaia performances (assuming the orbital photocentre shift is measured with a similar accuracy than the parallaxes).

Accounting for 5\% uncertainty on the total mass (shared between the interferometric resolved astrometry and the Gaia distance, see Eq.~\ref{eq:k3}), and for 1 to 10\% uncertainty on the mass-ratio (shared between the interferometric flux ratio and the photocentre shift amplitude from Gaia, see Eq.~\ref{eq:qFromGaia}), we expect to measure the individual masses with an uncertainty of $5-15$\%, independently of radial velocities.

\section{Conclusion}
\label{sec:conclu}

In this paper we report $\approx$$130$ observations of spatially resolved astrometry on ten massive O-type binaries, using OLBI. This dataset is used to derive the elements of the apparent orbits. Combined with the distance, it provides the total mass of the system. We also compute preliminary individual component masses for the four systems with available SB2 radial velocities. The conclusions are the following:
\begin{itemize}
\item Nine over ten binaries have their three-dimensional orbits well constrained. One requires additional OLBI observations to lift remaining degeneracies.
\item We provide orbital elements for four colliding wind, non-thermal radio emitters. They constitute valuable targets for future high-angular resolution radio imaging.
\item The large range of eccentricity suggests that the current lack of long-period systems with low ($e<0.25$) or high ($e>0.75$) eccentricities results from observational biases.
\item For the most-studied systems, we find a clear conflict between the astrometric orbit and the SB2 radial velocity amplitudes. We speculate that the SB2 amplitudes are affected by systematic biases, that could represent an intrinsic limitation for estimating dynamical masses from OLBI+SB2 or Gaia+SB2.
\item Our results can be combined with future Gaia parallaxes and orbital photocentre displacements to measure the masses of the individual components with an accuracy of 5 to 15\%, independently of the radial velocities.
\end{itemize}

\begin{acknowledgements} 
PIONIER is funded by the Universit\'e Joseph Fourier (now Universit\'e Grenoble Alpes, UGA), the Institut de Plan\'etologie et d'Astrophysique de Grenoble (IPAG), the Agence Nationale pour la Recherche (ANR-06-BLAN-0421, ANR-10-BLAN-0505, ANR-10-LABX56, ANR-11-LABX-13), and the Institut National des Sciences de l'Univers (INSU PNP and PNPS). Its integrated optics beam combiner is the result of a collaboration between IPAG and CEA-LETI based on CNES R\&T funding. This work is based on observations made with the ESO telescopes. It made use of the Smithsonian/NASA Astrophysics Data System (ADS) and of the Centre de Donn\'ees astronomiques de Strasbourg (CDS). All calculations and graphics were performed with the freeware \texttt{Yorick}. Some of the observations were acquired in the framework of the Belgium Guaranteed Time of VLTI (VISA project). The authors warmly thank everyone involved in the VLTI project. The authors warmly thank the entire GRAVITY consortium. GRAVITY is developed in a collaboration by the Max Planck Institute for Extraterrestrial Physics, LESIA of Paris Observatory and IPAG of Universit\'e Grenoble Alpes/CNRS, Max Planck Institute for Astronomy, the University of Cologne, the Centro Multidisciplinar de Astrofísica Lisbon and Porto, and the European Southern Observatory
\end{acknowledgements}

\bibliographystyle{/Users/lebouquj/Tex/AandA/aa}  
\bibliography{/Users/lebouquj/Biblio/BibTex/allNew}   

\listofobjects{}

\appendix
\section{Individual results}
\label{sec:individualresults}

\begin{figure*}  \centering 
  \includegraphics[scale=1]{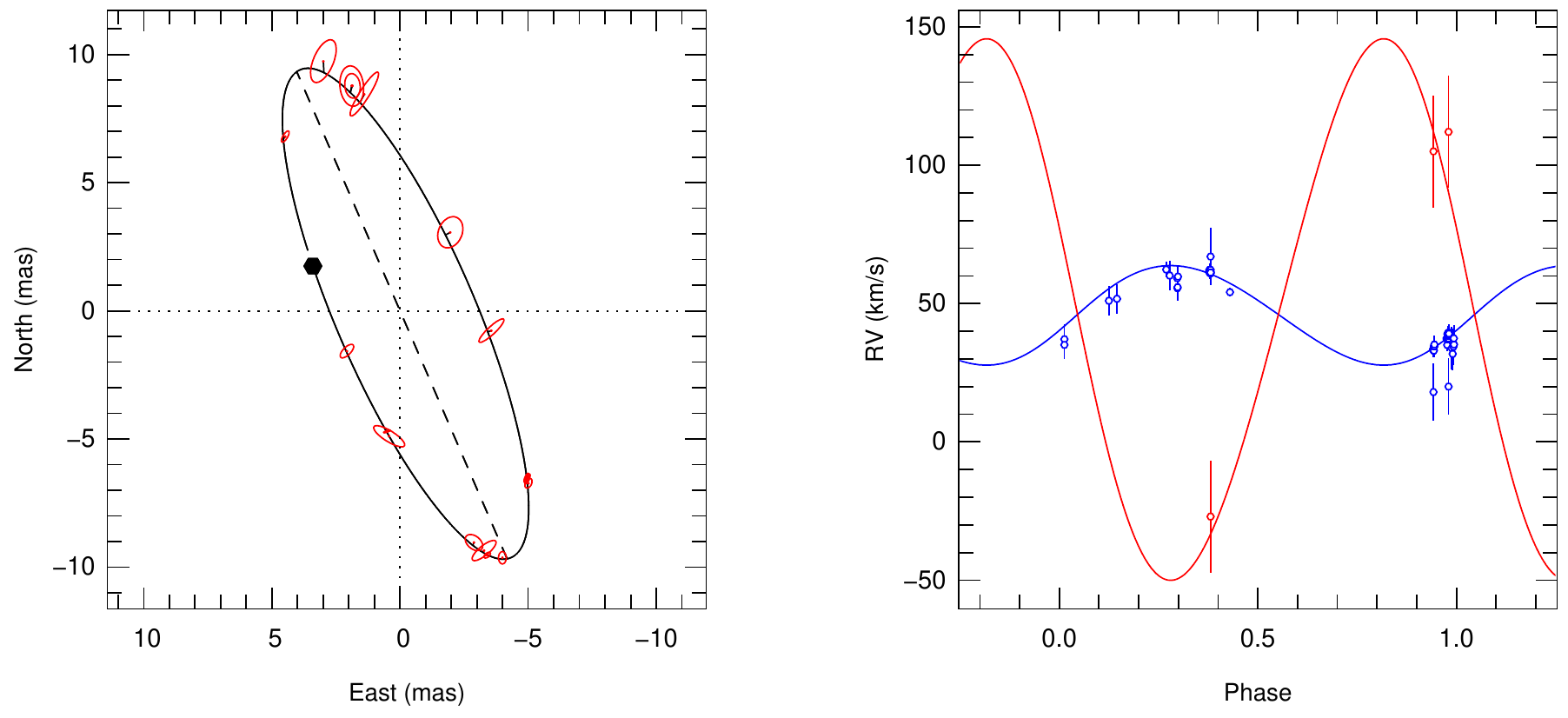}
\end{figure*}
\begin{figure*}  \centering 
  \begin{minipage}[c]{9cm} \centering \includegraphics[scale=1]{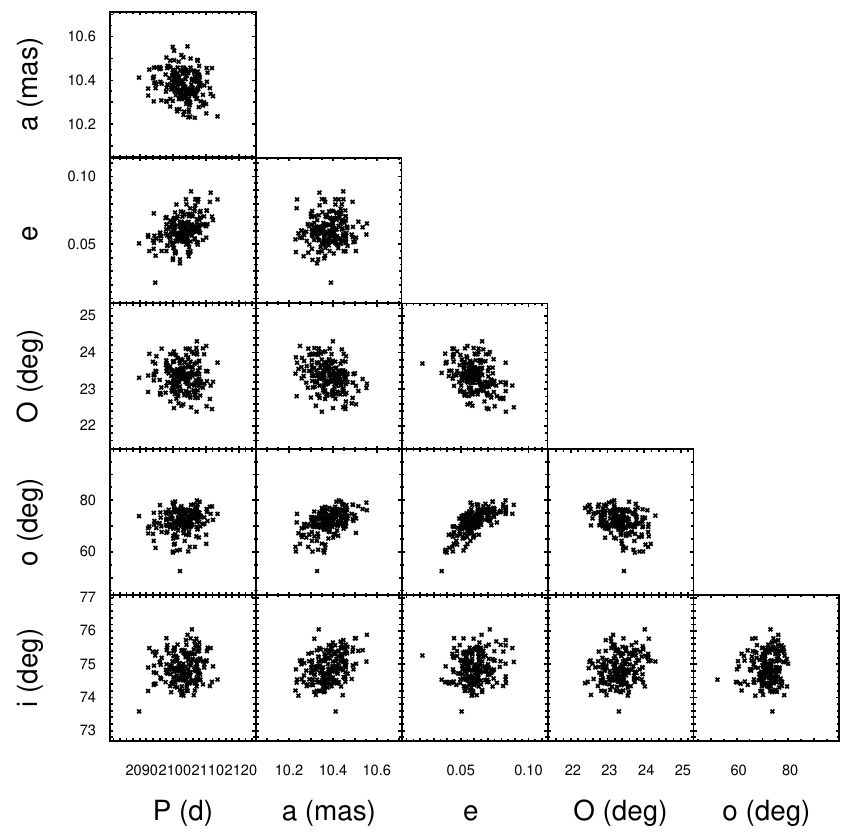} \end{minipage} \hfill
  \begin{minipage}[c]{8cm}\centering  \begin{tabular*}{6.35cm}{cccc}
 \hline\hline\noalign{\smallskip}
 Element & Unit & Value & Uncertainty \\
 \noalign{\smallskip}\hline\noalign{\smallskip}
 $T$ & MJD  &  $54045$  &  $32$  \\
 $P$ & days  &  $2103.3$  &  $4.3$  \\
 $a$ & mas  &  $10.383$  &  $0.065$  \\
 $e$ &   &  $0.060$  &  $0.010$  \\
 $\Omega$ & deg  &  $23.36$  &  $0.39$  \\
 $\omega$ & deg  &  $72.4$  &  $5.5$  \\
 $i$ & deg  &  $74.87$  &  $0.43$  \\
 $f_H$ &   &  $0.78$  &  $0.02$  \\
 $K_a$ & km/s  &  $17.97$  &  $0.79$  \\
 $K_b$ & km/s  &  $98$  &  $18$  \\
 $g$ & km/s  &  $46.05$  &  $0.28$  \\
 \noalign{\smallskip}\hline\noalign{\smallskip}
 \multicolumn{4}{c}{From apparent orbit and distance}\\
 $d$ & pc  &  $1100$  &  $100$  \\
 $M_t$ & ${M}_\odot$  &  $45$  &  $12$  \\
 \noalign{\smallskip}\hline\noalign{\smallskip}
 \multicolumn{4}{c}{From apparent orbit and radial velocities}\\
 $d$ & pc  &  $2229$  &  $345$  \\
 $M_a$ & ${M}_\odot$  &  $316$  &  $168$  \\
 $M_b$ & ${M}_\odot$  &  $58$  &  $19$  \\
 \noalign{\smallskip}\hline
 \end{tabular*}
 \end{minipage}
  \caption{Best fit orbital solution to the astrometric and velocimetric observations of HD54662. Top-left: motion of the secondary around the primary. The periastron of the secondary is represented by a filled symbol and the line of nodes by a dashed line. Top-right: radial velocities of the primary (blue) and the secondary (red). Bottom-left: best fit parameters with random noise on the dataset. Bottom-right: best fit parameters and uncertainties.}
\end{figure*}

\clearpage
\begin{figure*}  \centering 
  \includegraphics[scale=1]{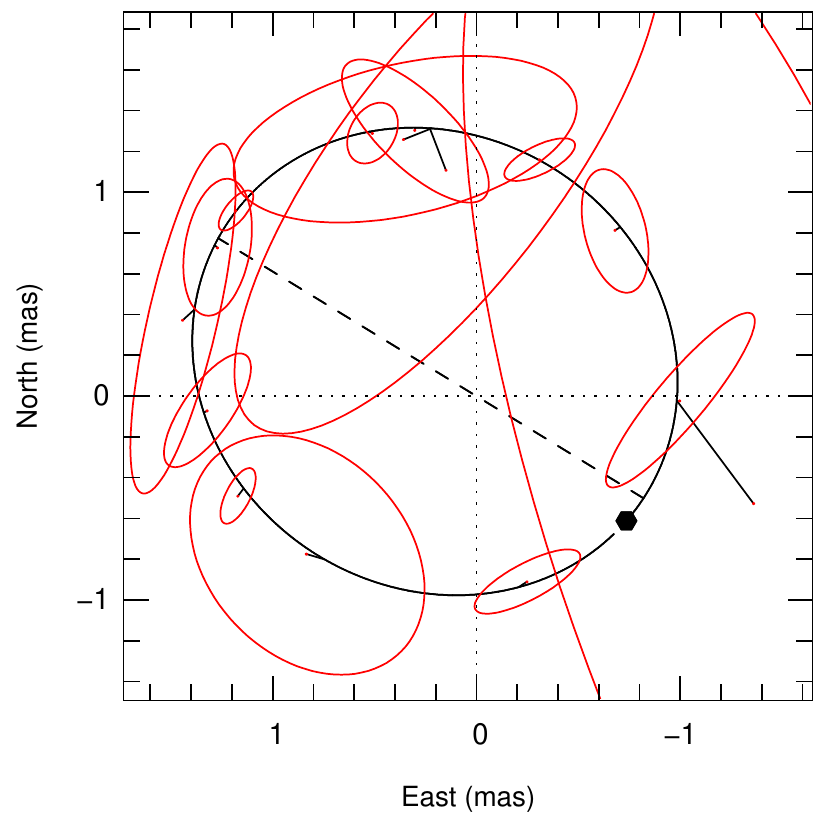}
\end{figure*}
\begin{figure*}  \centering 
  \begin{minipage}[c]{9cm} \centering \includegraphics[scale=1]{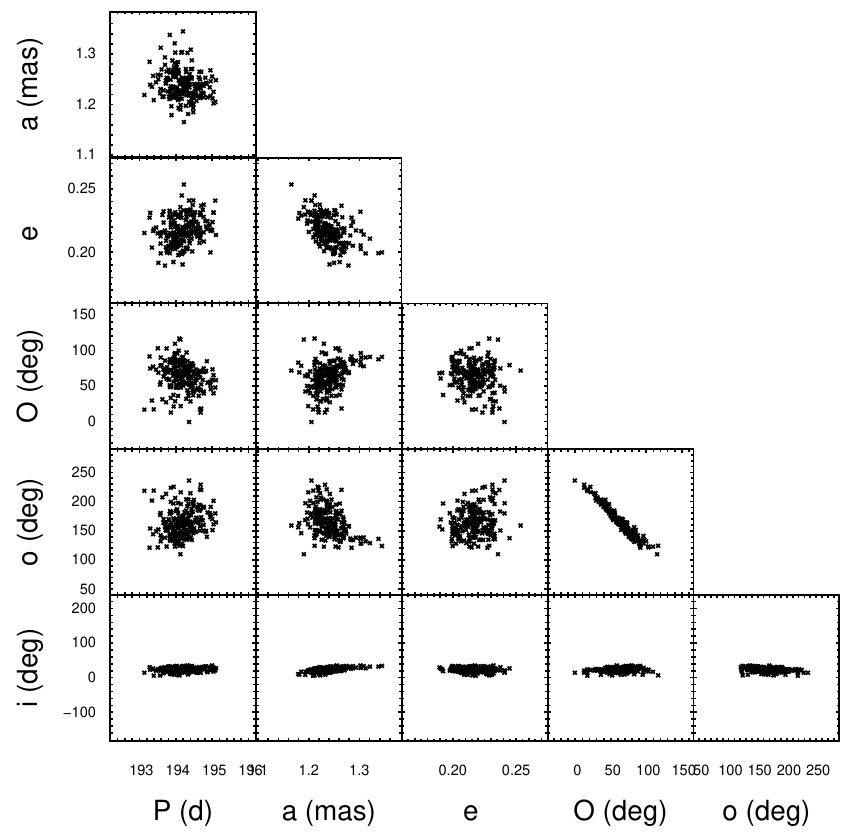} \end{minipage} \hfill
  \begin{minipage}[c]{8cm}\centering  \begin{tabular*}{6.35cm}{cccc}
 \hline\hline\noalign{\smallskip}
 Element & Unit & Value & Uncertainty \\
 \noalign{\smallskip}\hline\noalign{\smallskip}
 $T$ & MJD  &  $54857.7$  &  $5.4$  \\
 $P$ & days  &  $194.31$  &  $0.39$  \\
 $a$ & mas  &  $1.224$  &  $0.028$  \\
 $e$ &   &  $0.217$  &  $0.011$  \\
 $\Omega$ & deg  &  $59$  &  $20$  \\
 $\omega$ & deg  &  $171$  &  $24$  \\
 $i$ & deg  &  $22$  &  $41$  \\
 $f_H$ &   &  $0.86$  &  $0.02$  \\
 \noalign{\smallskip}\hline\noalign{\smallskip}
 \multicolumn{4}{c}{From apparent orbit and distance}\\
 $d$ & pc  &  $2350$  &  $200$  \\
 $M_t$ & ${M}_\odot$  &  $84$  &  $22$  \\
 \noalign{\smallskip}\hline
 \end{tabular*}
 \end{minipage}
  \caption{Best fit orbital solution to the astrometric observations of HD93250. Top: motion of the secondary around the primary. The periastron of the secondary is represented by a filled symbol and the line of nodes by a dashed line. Bottom-left: best fit parameters with random noise on the dataset. Bottom-right: best fit parameters and uncertainties.}
\end{figure*}

\clearpage
\begin{figure*}  \centering 
  \includegraphics[scale=1]{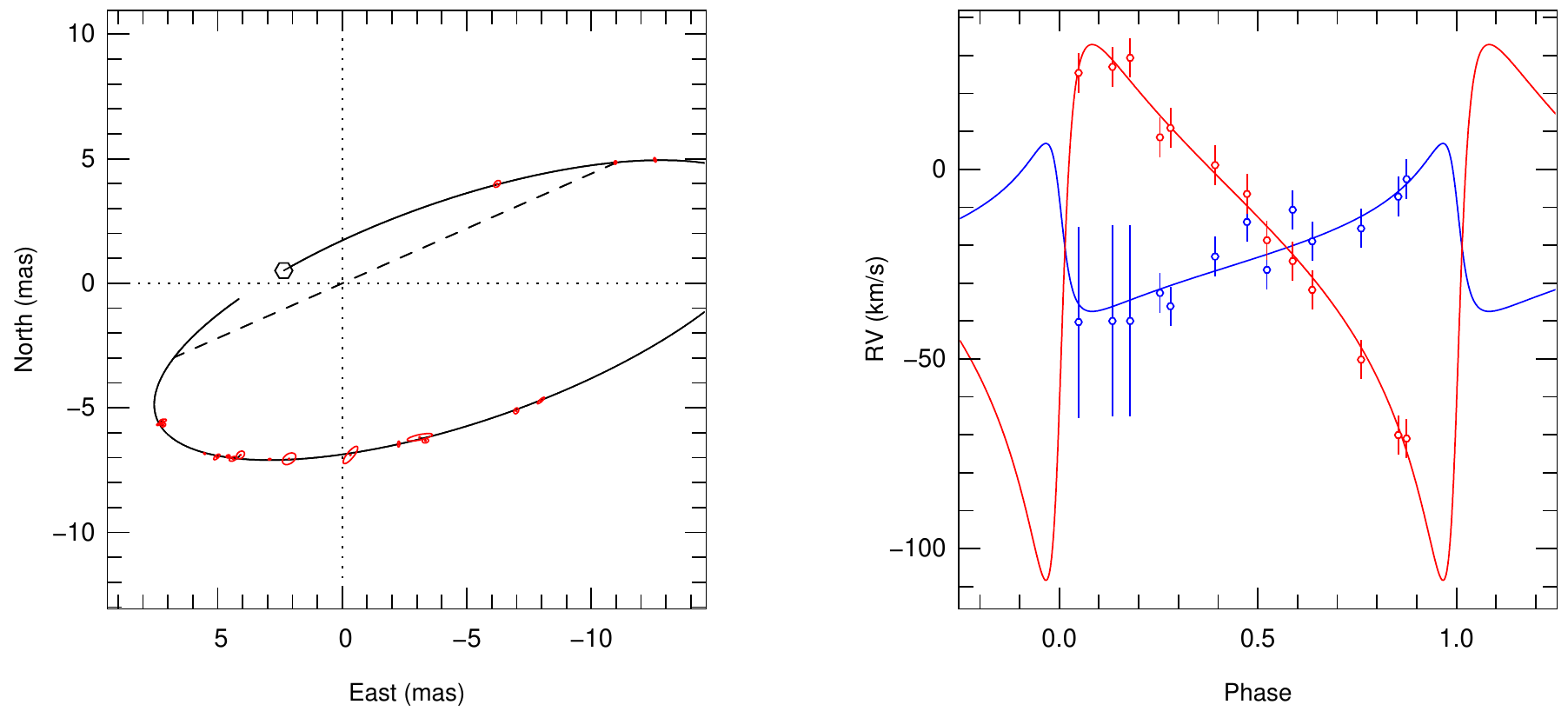}
\end{figure*}
\begin{figure*}  \centering 
  \begin{minipage}[c]{9cm} \centering \includegraphics[scale=1]{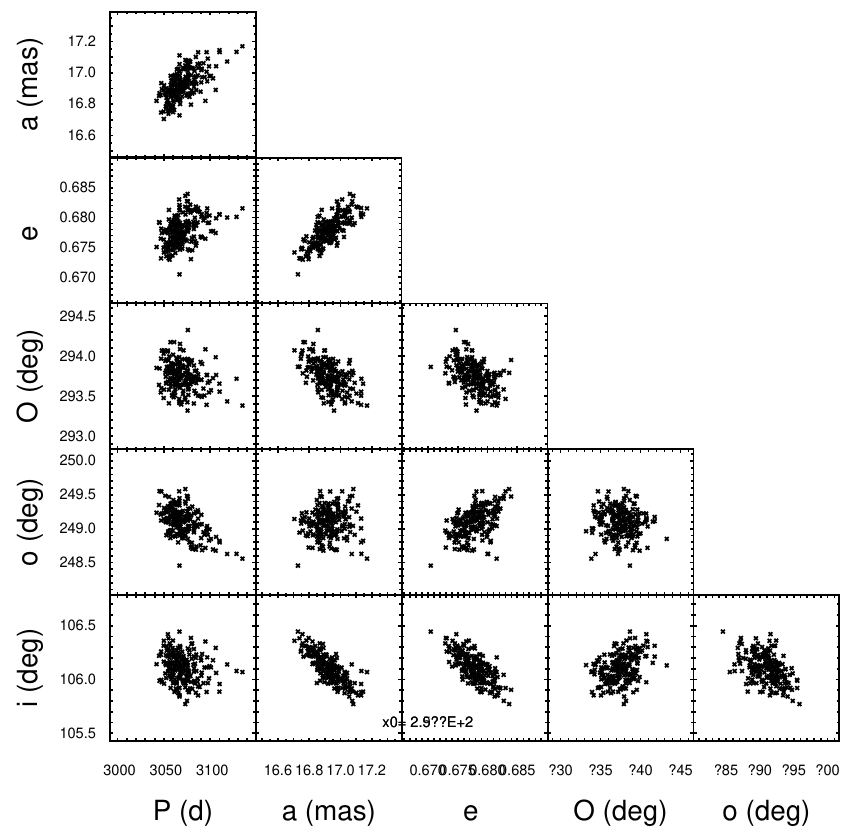} \end{minipage} \hfill
  \begin{minipage}[c]{8cm}\centering  \begin{tabular*}{6.35cm}{cccc}
 \hline\hline\noalign{\smallskip}
 Element & Unit & Value & Uncertainty \\
 \noalign{\smallskip}\hline\noalign{\smallskip}
 $T$ & MJD  &  $51179$  &  $30$  \\
 $P$ & days  &  $3069$  &  $15$  \\
 $a$ & mas  &  $16.918$  &  $0.091$  \\
 $e$ &   &  $0.6780$  &  $0.0024$  \\
 $\Omega$ & deg  &  $293.75$  &  $0.18$  \\
 $\omega$ & deg  &  $249.12$  &  $0.21$  \\
 $i$ & deg  &  $106.11$  &  $0.13$  \\
 $f_H$ &   &  $0.24$  &  $0.02$  \\
 $K_a$ & km/s  &  $22.1$  &  $3.7$  \\
 $K_b$ & km/s  &  $70.6$  &  $3.2$  \\
 $g$ & km/s  &  $-20.7$  &  $1.1$  \\
 \noalign{\smallskip}\hline\noalign{\smallskip}
 \multicolumn{4}{c}{From apparent orbit and distance}\\
 $d$ & pc  &  $1320$  &  $120$  \\
 $M_t$ & ${M}_\odot$  &  $158$  &  $43$  \\
 \noalign{\smallskip}\hline\noalign{\smallskip}
 \multicolumn{4}{c}{From apparent orbit and radial velocities}\\
 $d$ & pc  &  $1184$  &  $62$  \\
 $M_a$ & ${M}_\odot$  &  $87$  &  $12$  \\
 $M_b$ & ${M}_\odot$  &  $27.1$  &  $7.0$  \\
 \noalign{\smallskip}\hline
 \end{tabular*}
 \end{minipage}
  \caption{ \label{fig:HD150136} Best fit orbital solution to the astrometric and velocimetric observations of HD150136.  Top-left: motion of the secondary around the primary. The periastron of the secondary is represented by an open symbol and the line of nodes by a dashed line. Top-right: radial velocities of the primary (blue) and the secondary (red). Bottom-left: best fit parameters with random noise on the dataset. Bottom-right: best fit parameters and uncertainties.}
\end{figure*}

\clearpage
\begin{figure*}  \centering 
  \includegraphics[scale=1]{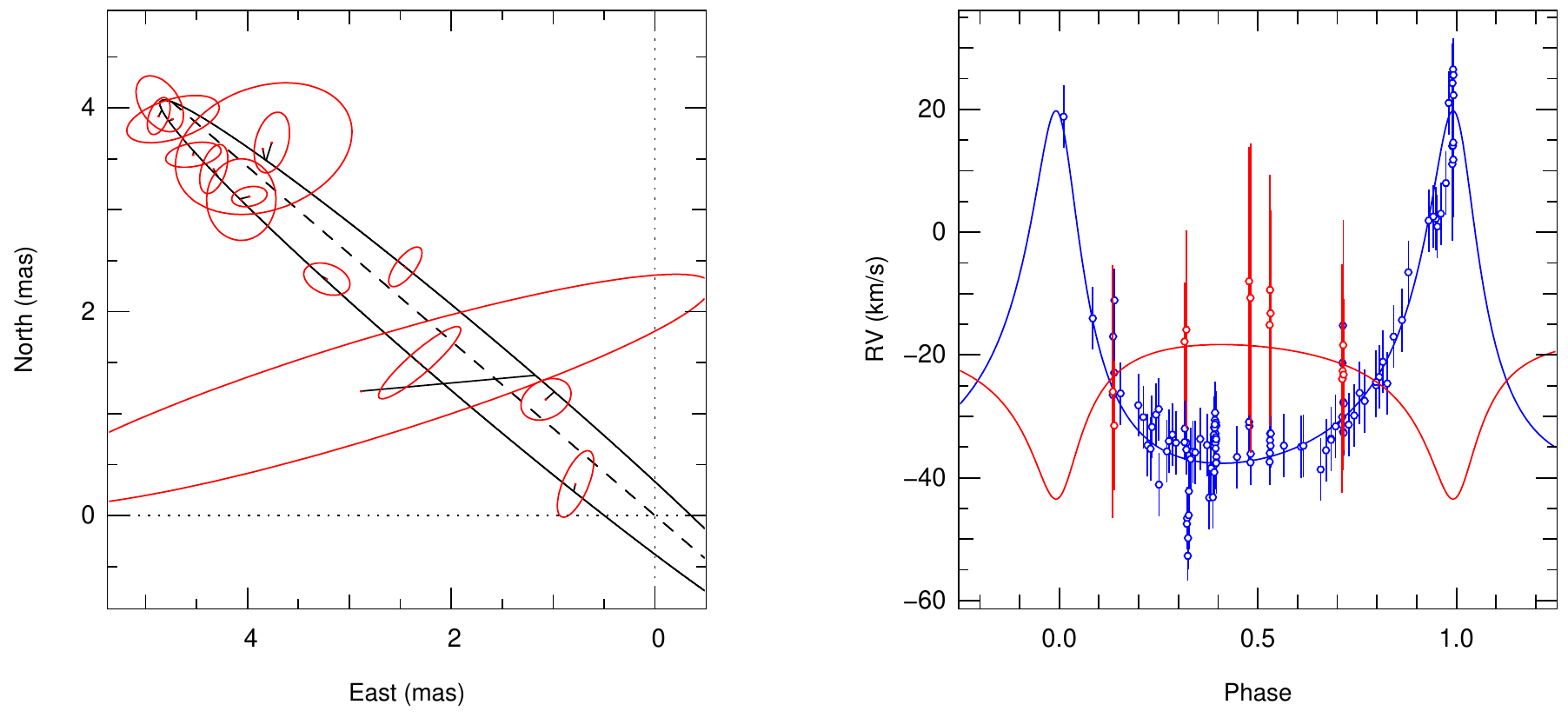}
\end{figure*}
\begin{figure*}  \centering 
  \begin{minipage}[c]{9cm} \centering \includegraphics[scale=1]{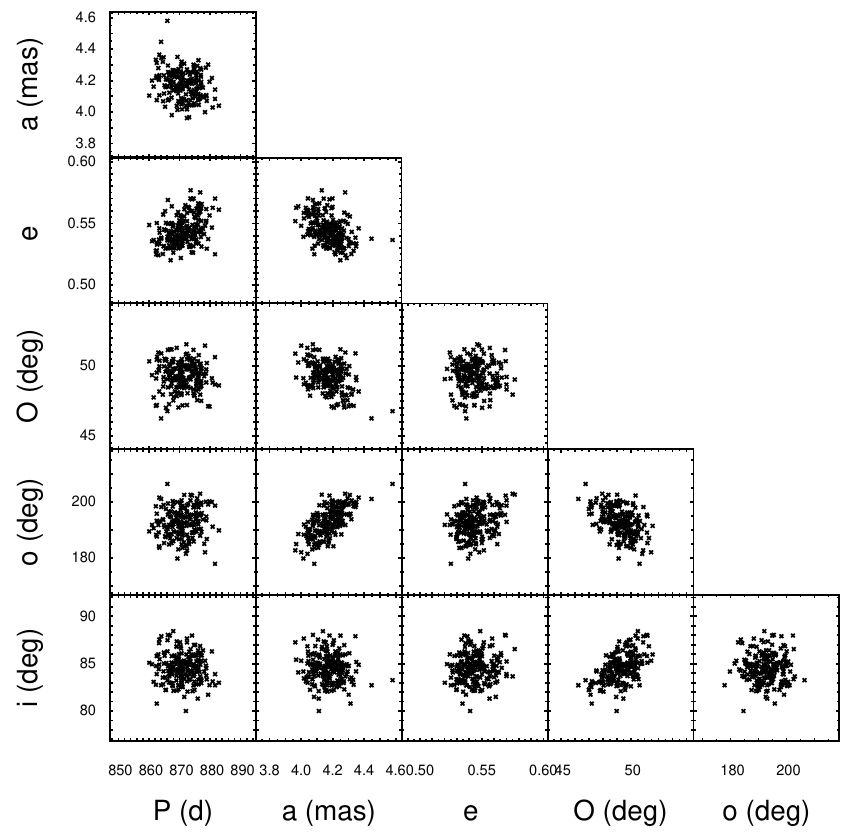} \end{minipage} \hfill
  \begin{minipage}[c]{8cm}\centering  \begin{tabular*}{6.35cm}{cccc}
 \hline\hline\noalign{\smallskip}
 Element & Unit & Value & Uncertainty \\
 \noalign{\smallskip}\hline\noalign{\smallskip}
 $T$ & MJD  &  $56276$  &  $11$  \\
 $P$ & days  &  $871.3$  &  $4.6$  \\
 $a$ & mas  &  $4.150$  &  $0.090$  \\
 $e$ &   &  $0.543$  &  $0.011$  \\
 $\Omega$ & deg  &  $49.4$  &  $1.0$  \\
 $\omega$ & deg  &  $192.0$  &  $5.1$  \\
 $i$ & deg  &  $84.7$  &  $1.5$  \\
 $f_H$ &   &  $0.15$  &  $0.02$  \\
 $K_a$ & km/s  &  $28.7$  &  $1.7$  \\
 $K_b$ & km/s  &  $12.6$  &  $9.6$  \\
 $g$ & km/s  &  $-24.20$  &  $0.81$  \\
 \noalign{\smallskip}\hline\noalign{\smallskip}
 \multicolumn{4}{c}{From apparent orbit and distance}\\
 $d$ & pc  &  $1523$  &  $100$  \\
 $M_t$ & ${M}_\odot$  &  $44.4$  &  $9.5$  \\
 \noalign{\smallskip}\hline\noalign{\smallskip}
 \multicolumn{4}{c}{From apparent orbit and radial velocities}\\
 $d$ & pc  &  $672$  &  $152$  \\
 $M_a$ & ${M}_\odot$  &  $1.2$  &  $2.0$  \\
 $M_b$ & ${M}_\odot$  &  $2.6$  &  $1.3$  \\
 \noalign{\smallskip}\hline
 \end{tabular*}
 \end{minipage}
  \caption{Best fit orbital solution to the astrometric and velocimetric observations of HD152233. Top-left: motion of the secondary around the primary. The line of nodes is represented by a dashed line. Top-right: radial velocities of the primary (blue) and the secondary (red).  Bottom-left: best fit parameters with random noise on the dataset. Bottom-right: best fit parameters and uncertainties.}
\end{figure*}

\clearpage
\begin{figure*}  \centering 
  \includegraphics[scale=1]{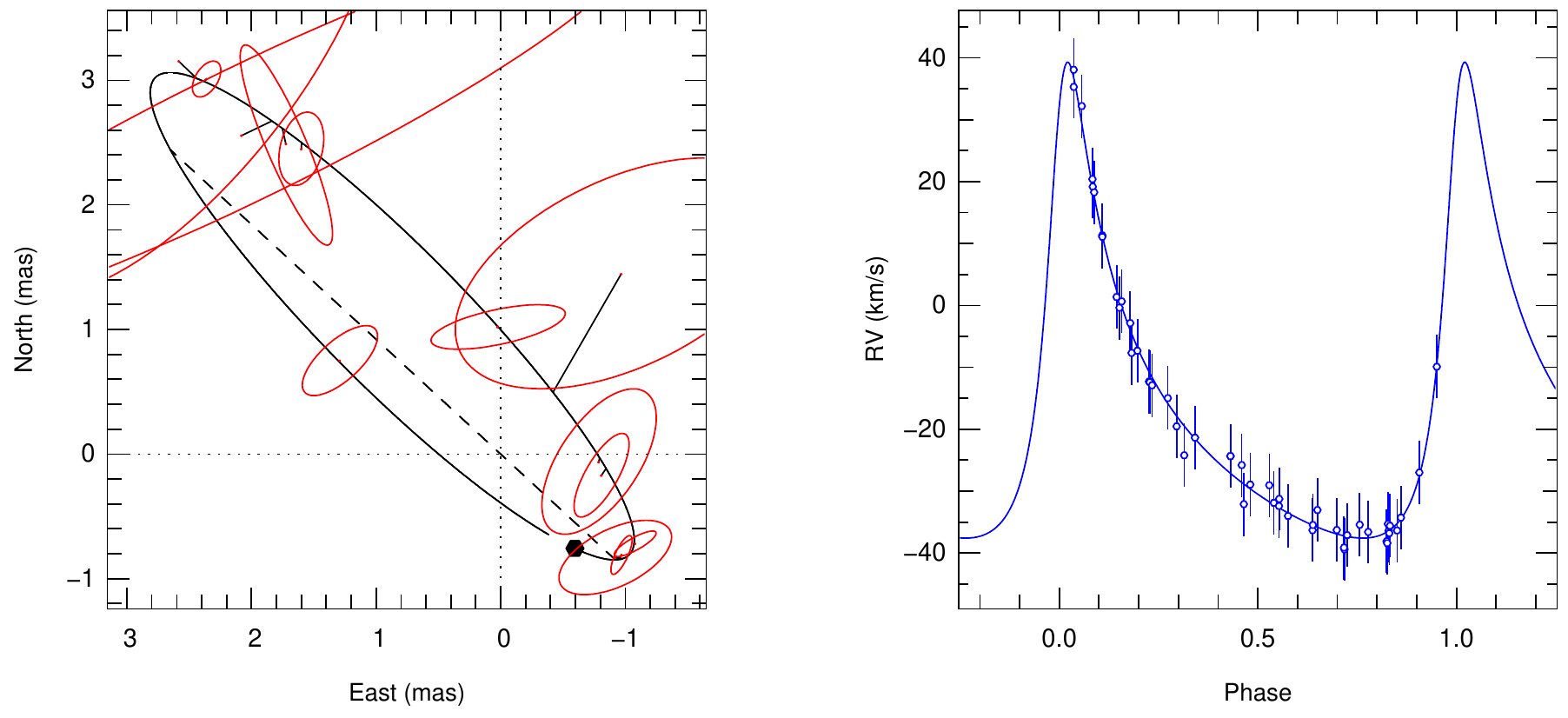}
\end{figure*}
\begin{figure*}  \centering 
  \begin{minipage}[c]{9cm} \centering \includegraphics[scale=1]{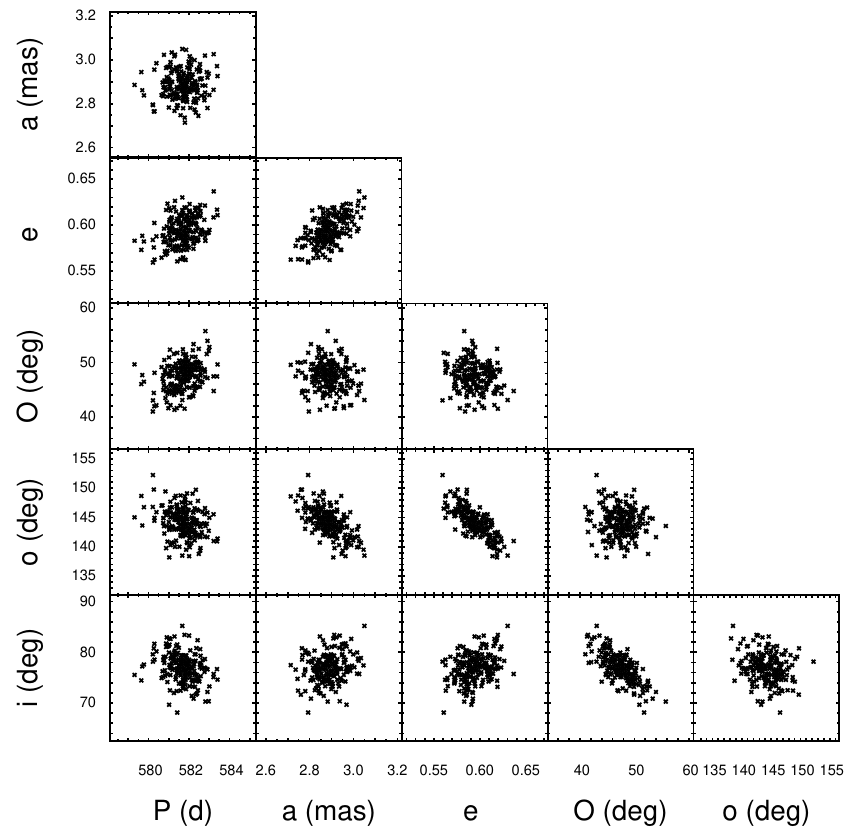} \end{minipage} \hfill
  \begin{minipage}[c]{8cm}\centering  \begin{tabular*}{6.35cm}{cccc}
 \hline\hline\noalign{\smallskip}
 Element & Unit & Value & Uncertainty \\
 \noalign{\smallskip}\hline\noalign{\smallskip}
 $T$ & MJD  &  $71266$  &  $19$  \\
 $P$ & days  &  $581.71$  &  $0.70$  \\
 $a$ & mas  &  $2.871$  &  $0.065$  \\
 $e$ &   &  $0.593$  &  $0.015$  \\
 $\Omega$ & deg  &  $47.4$  &  $2.6$  \\
 $\omega$ & deg  &  $144.6$  &  $2.4$  \\
 $i$ & deg  &  $76.8$  &  $2.8$  \\
 $f_H$ &   &  $0.27$  &  $0.02$  \\
 $K_a$ & km/s  &  $38.4$  &  $1.6$  \\
 $g$ & km/s  &  $-17.75$  &  $0.83$  \\
 \noalign{\smallskip}\hline\noalign{\smallskip}
 \multicolumn{4}{c}{From apparent orbit and distance}\\
 $d$ & pc  &  $1523$  &  $100$  \\
 $M_t$ & ${M}_\odot$  &  $32.9$  &  $6.8$  \\
 \noalign{\smallskip}\hline
 \end{tabular*}
 \end{minipage}
  \caption{Best fit orbital solution to the astrometric observations of HD152247. Top-left: motion of the secondary around the primary. The periastron of the secondary is represented by a filled symbol and the line of nodes by a dashed line. Top-right: SB1 radial velocities. Bottom-left: best fit parameters with random noise on the dataset. Bottom-right: best fit parameters and uncertainties.}
\end{figure*}

\clearpage
\begin{figure*}  \centering 
  \includegraphics[scale=1]{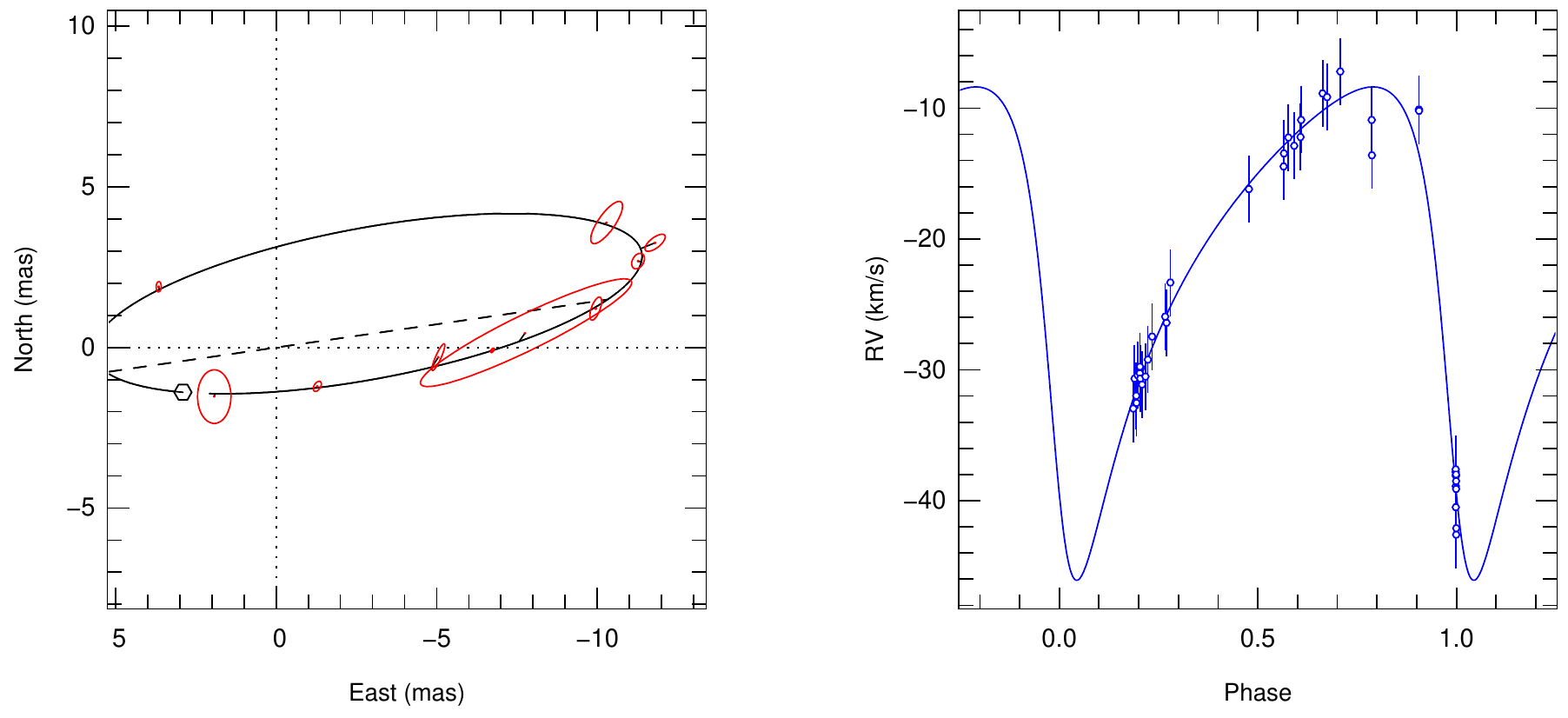}
\end{figure*}
\begin{figure*}  \centering 
  \begin{minipage}[c]{9cm} \centering \includegraphics[scale=1]{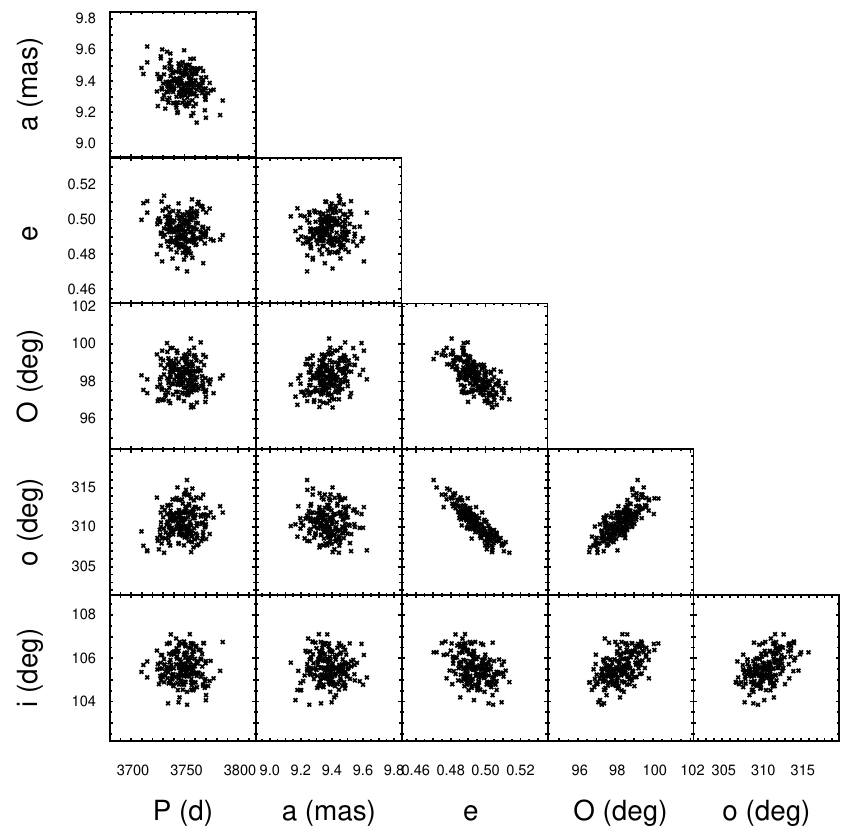} \end{minipage} \hfill
  \begin{minipage}[c]{8cm}\centering  \begin{tabular*}{6.35cm}{cccc}
 \hline\hline\noalign{\smallskip}
 Element & Unit & Value & Uncertainty \\
 \noalign{\smallskip}\hline\noalign{\smallskip}
 $T$ & MJD  &  $56886$  &  $12$  \\
 $P$ & days  &  $3749$  &  $13$  \\
 $a$ & mas  &  $9.365$  &  $0.091$  \\
 $e$ &   &  $0.4948$  &  $0.0081$  \\
 $\Omega$ & deg  &  $98.23$  &  $0.76$  \\
 $\omega$ & deg  &  $310.5$  &  $1.8$  \\
 $i$ & deg  &  $105.55$  &  $0.66$  \\
 $f_H$ &   &  $0.36$  &  $0.02$  \\
 $K_a$ & km/s  &  $18.85$  &  $0.86$  \\
 $g$ & km/s  &  $-21.17$  &  $0.45$  \\
 \noalign{\smallskip}\hline\noalign{\smallskip}
 \multicolumn{4}{c}{From apparent orbit and distance}\\
 $d$ & pc  &  $1523$  &  $100$  \\
 $M_t$ & ${M}_\odot$  &  $27.5$  &  $5.4$  \\
 \noalign{\smallskip}\hline
 \end{tabular*}
 \end{minipage}
  \caption{\label{fig:HD152314}  Best fit orbital solution to the astrometric observations of HD152314. Top-left: motion of the secondary around the primary. The periastron of the secondary is represented by an open symbol and the line of nodes by a dashed line. Top-right: SB1 radial velocities. Bottom-left: best fit parameters with random noise on the dataset. Bottom-right: best fit parameters and uncertainties.}
\end{figure*}

\clearpage
\begin{figure*}  \centering 
  \includegraphics[scale=1]{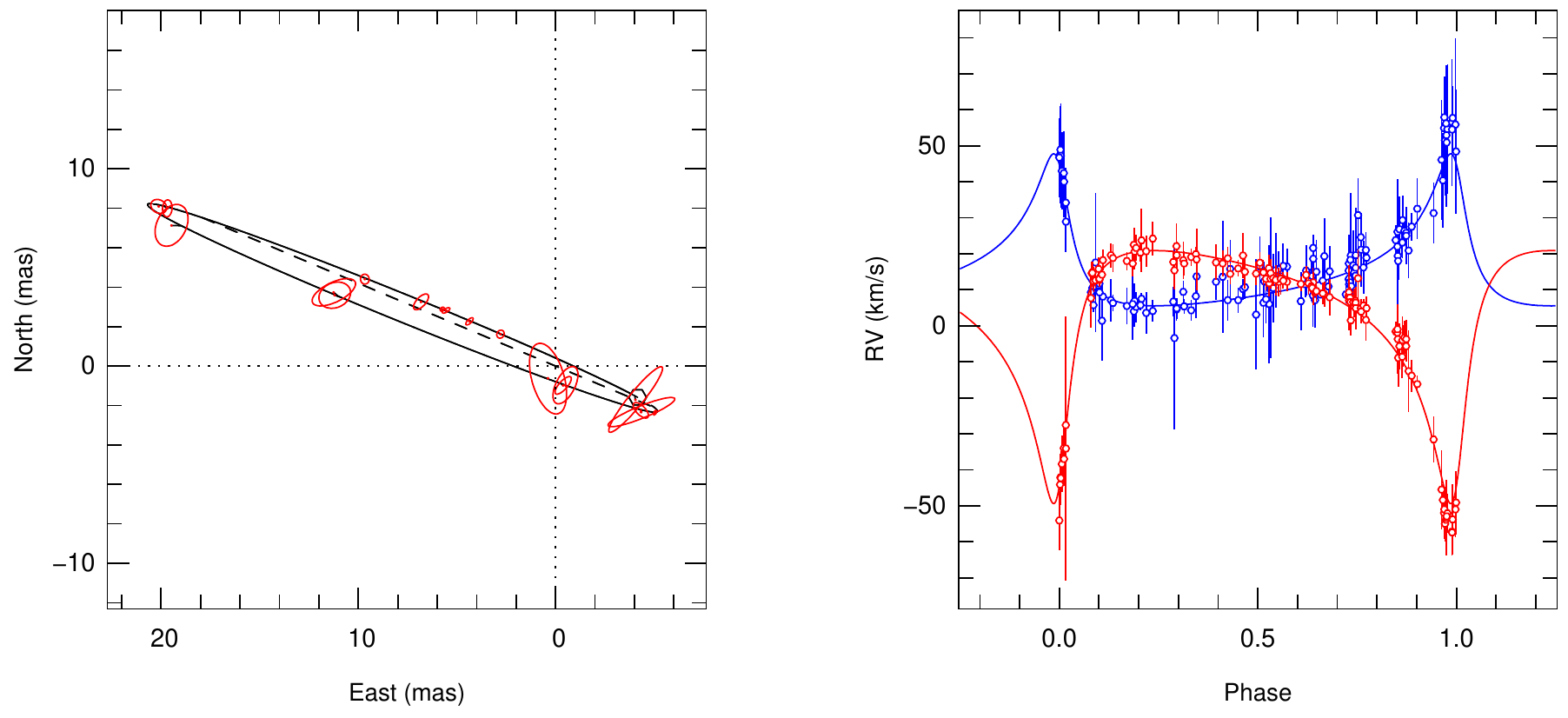}
\label{fig:HD164794}
\end{figure*}
\begin{figure*}  \centering 
  \begin{minipage}[c]{9cm} \centering \includegraphics[scale=1]{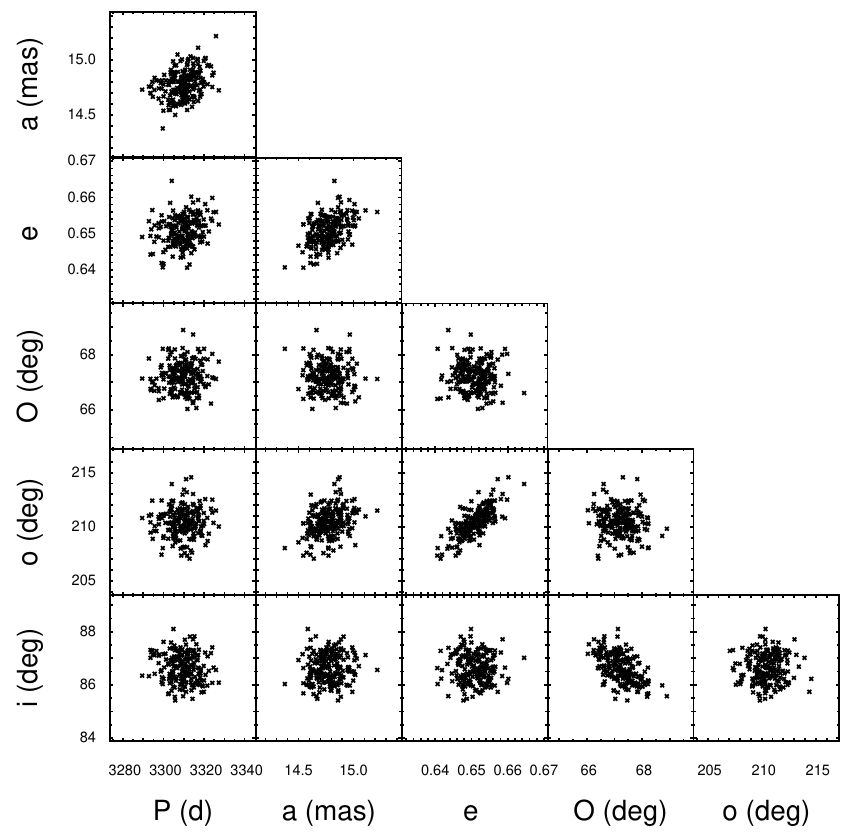} \end{minipage} \hfill
  \begin{minipage}[c]{8cm}\centering  \begin{tabular*}{6.35cm}{cccc}
 \hline\hline\noalign{\smallskip}
 Element & Unit & Value & Uncertainty \\
 \noalign{\smallskip}\hline\noalign{\smallskip}
 $T$ & MJD  &  $46613$  &  $21$  \\
 $P$ & days  &  $3310.4$  &  $7.0$  \\
 $a$ & mas  &  $14.78$  &  $0.13$  \\
 $e$ &   &  $0.6508$  &  $0.0039$  \\
 $\Omega$ & deg  &  $67.21$  &  $0.51$  \\
 $\omega$ & deg  &  $210.4$  &  $1.3$  \\
 $i$ & deg  &  $86.64$  &  $0.53$  \\
 $f_H$ &   &  $0.62$  &  $0.02$  \\
 $K_a$ & km/s  &  $21.1$  &  $1.0$  \\
 $K_b$ & km/s  &  $35.14$  &  $0.74$  \\
 $g$ & km/s  &  $14.79$  &  $0.29$  \\
 \noalign{\smallskip}\hline\noalign{\smallskip}
 \multicolumn{4}{c}{From apparent orbit and distance}\\
 $d$ & pc  &  $1250$  &  $100$  \\
 $M_t$ & ${M}_\odot$  &  $77$  &  $18$  \\
 \noalign{\smallskip}\hline\noalign{\smallskip}
 \multicolumn{4}{c}{From apparent orbit and radial velocities}\\
 $d$ & pc  &  $881$  &  $22$  \\
 $M_a$ & ${M}_\odot$  &  $16.77$  &  $0.99$  \\
 $M_b$ & ${M}_\odot$  &  $10.07$  &  $0.89$  \\
 \noalign{\smallskip}\hline
 \end{tabular*}
 \end{minipage}
  \caption{Best fit orbital solution to the astrometric and velocimetric observations of HD164794. Top-left: motion of the secondary around the primary. The periastron of the secondary is represented by an open symbol and the line of nodes by a dashed line. Top-right: radial velocities of the primary (blue) and the secondary (red). Bottom-left: best fit parameters with random noise on the dataset. Bottom-right: best fit parameters and uncertainties.}
\end{figure*}

\clearpage
\begin{figure*}  \centering 
  \includegraphics[scale=1]{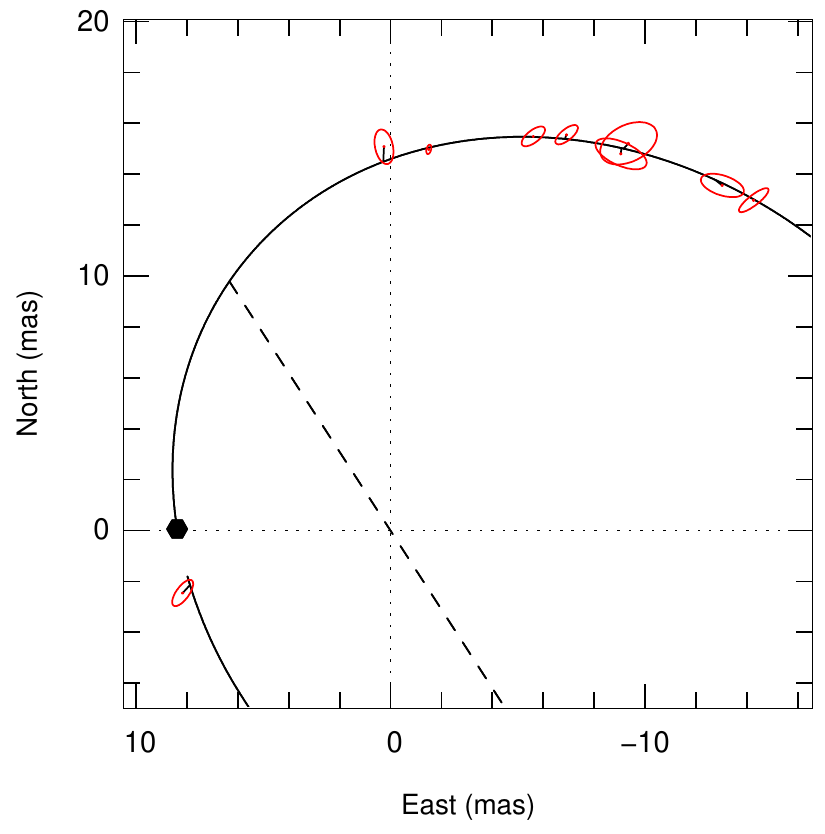}
\end{figure*}
\begin{figure*}  \centering 
  \begin{minipage}[c]{9cm} \centering \includegraphics[scale=1]{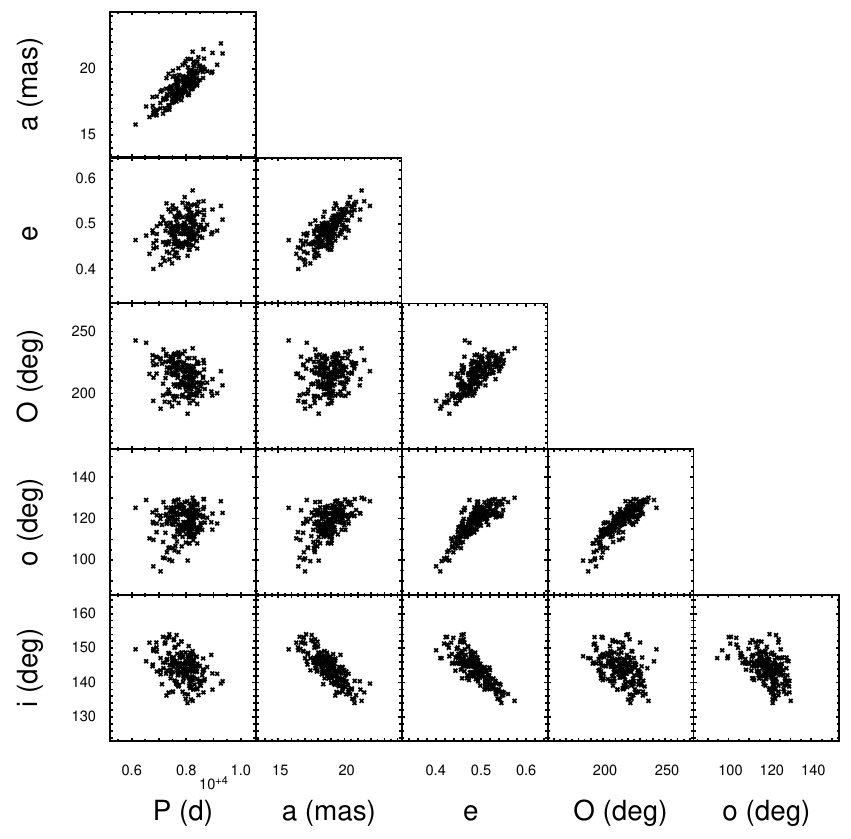} \end{minipage} \hfill
  \begin{minipage}[c]{8cm} \centering  \begin{tabular*}{6.35cm}{cccc}
 \hline\hline\noalign{\smallskip}
 Element & Unit & Value & Uncertainty \\
 \noalign{\smallskip}\hline\noalign{\smallskip}
 $T$ & MJD  &  $54696$  &  $50$  \\
 $P$ & days  &  $7895$  &  $520$  \\
 $a$ & mas  &  $18.6$  &  $1.1$  \\
 $e$ &   &  $0.479$  &  $0.031$  \\
 $\Omega$ & deg  &  $213$  &  $11$  \\
 $\omega$ & deg  &  $118.4$  &  $6.8$  \\
 $i$ & deg  &  $145.4$  &  $4.1$  \\
 $f_H$ &   &  $0.78$  &  $0.07$  \\
 \noalign{\smallskip}\hline\noalign{\smallskip}
 \multicolumn{4}{c}{From apparent orbit and distance}\\
 $d$ & pc  &  $1750$  &  $200$  \\
 $M_t$ & ${M}_\odot$  &  $74$  &  $29$  \\
 \noalign{\smallskip}\hline
 \end{tabular*}
 \end{minipage}
  \caption{\label{fig:HD167971} Best fit orbital solution to the astrometric observations of HD167971. Top: motion of the secondary around the primary. The periastron of the secondary is represented by a filled symbol and the line of nodes by a dashed line.  Bottom-left: best fit parameters with random noise on the dataset. Bottom-right: best fit parameters and uncertainties. This fit considers the constraints ($P=21.7\pm0.7\,$years, $e=0.53\pm0.05$) obtained by \citet{2013MNRAS.436..750I} from the times of minimum light of the inner pair.}
\end{figure*}

\clearpage
\begin{figure*}  \centering 
  \includegraphics[scale=1]{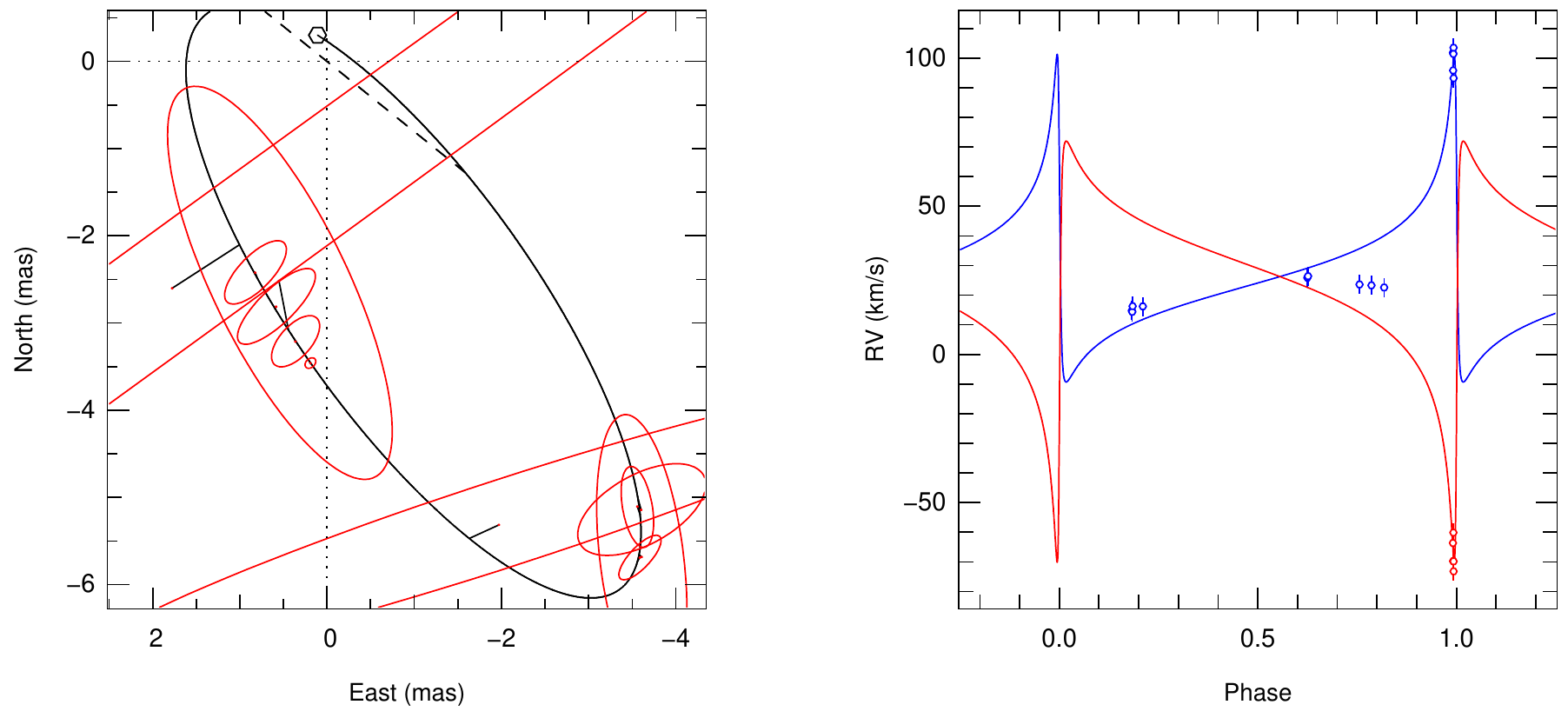}
\end{figure*}
\begin{figure*}  \centering 
  \begin{minipage}[c]{9cm} \centering \includegraphics[scale=1]{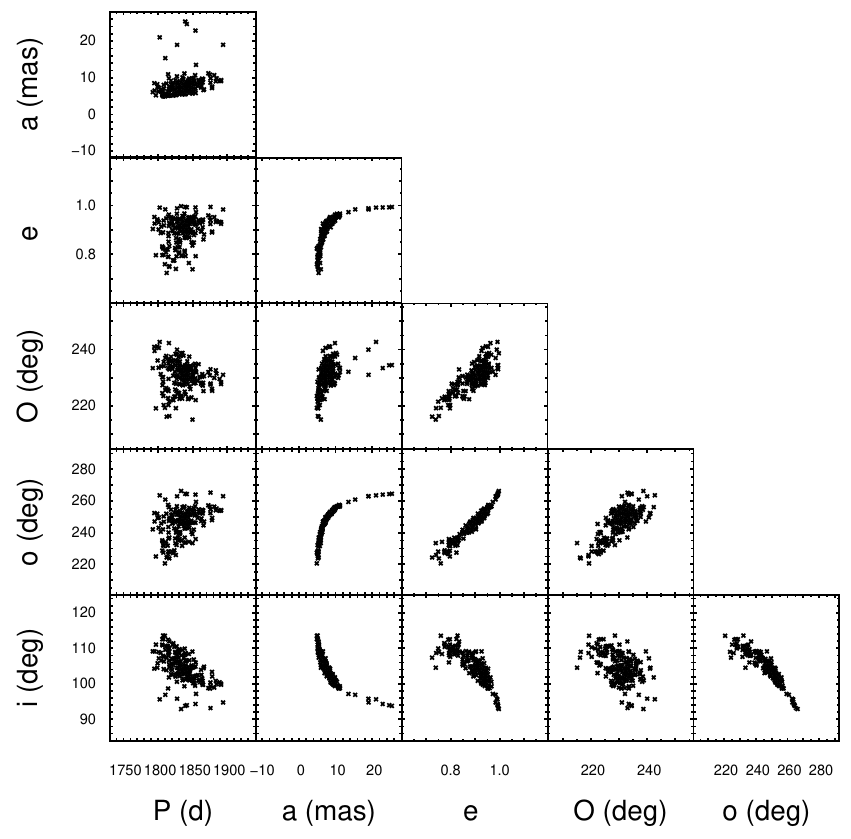} \end{minipage} \hfill
  \begin{minipage}[c]{8cm}\centering  \begin{tabular*}{6.35cm}{cccc}
 \hline\hline\noalign{\smallskip}
 Element & Unit & Value & Uncertainty \\
 \noalign{\smallskip}\hline\noalign{\smallskip}
 $T$ & MJD  &  $53525$  &  $14$  \\
 $P$ & days  &  $1836$  &  $20$  \\
 $a$ & mas  &  $7.0$  &  $3.9$  \\
 $e$ &   &  $0.902$  &  $0.058$  \\
 $\Omega$ & deg  &  $231.2$  &  $5.0$  \\
 $\omega$ & deg  &  $246.7$  &  $9.0$  \\
 $i$ & deg  &  $105.2$  &  $4.0$  \\
 $f_H$ &   &  $0.77$  &  $0.02$  \\
 $K_a$ & km/s  &  $55$  &  $35$  \\
 $K_b$ & km/s  &  $71$  &  $48$  \\
 $g$ & km/s  &  $26.3$  &  $9.2$  \\
 \noalign{\smallskip}\hline\noalign{\smallskip}
 \multicolumn{4}{c}{From apparent orbit and distance}\\
 $d$ & pc  &  $1800$  &  $100$  \\
 $M_t$ & ${M}_\odot$  &  $79$  &  $1004$  \\
 \noalign{\smallskip}\hline\noalign{\smallskip}
 \multicolumn{4}{c}{From apparent orbit and radial velocities}\\
 $d$ & pc  &  $1365$  &  $563$  \\
 $M_a$ & ${M}_\odot$  &  $19$  &  $26$  \\
 $M_b$ & ${M}_\odot$  &  $15$  &  $17$  \\
 \noalign{\smallskip}\hline
 \end{tabular*}
 \end{minipage}
  \caption{Best fit orbital solution to the astrometric and velocimetric observations of HD168137. Top-left: motion of the secondary around the primary. The periastron of the secondary is represented by an open symbol and the line of nodes by a dashed line. Top-right: radial velocities of the primary (blue) and the secondary (red). Bottom-left: best fit parameters with random noise on the dataset. Bottom-right: best fit parameters and uncertainties.}
\end{figure*}

\clearpage
\begin{figure*}  \centering 
  \includegraphics[scale=1]{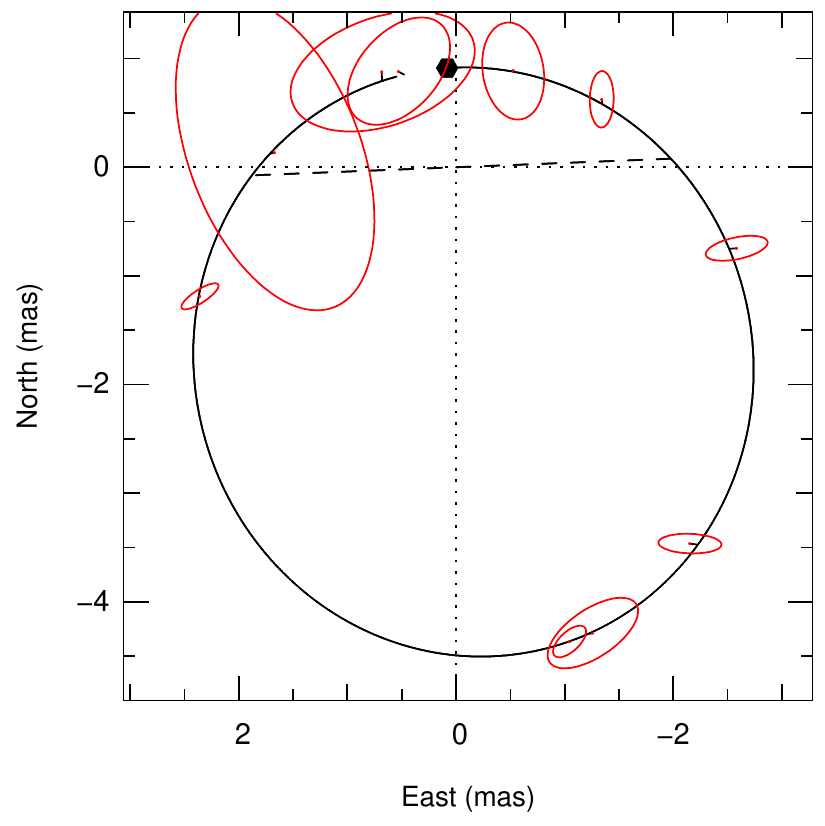}
\end{figure*}
\begin{figure*}  \centering 
  \begin{minipage}[c]{9cm} \centering \includegraphics[scale=1]{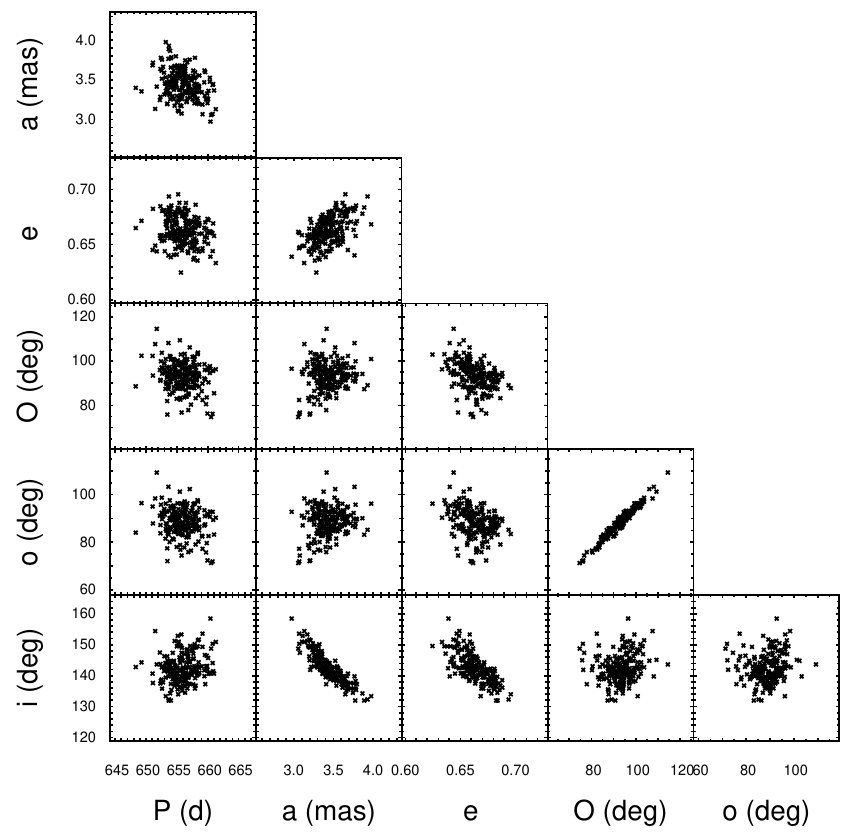} \end{minipage} \hfill
  \begin{minipage}[c]{8cm} \centering  \begin{tabular*}{6.35cm}{cccc}
 \hline\hline\noalign{\smallskip}
 Element & Unit & Value & Uncertainty \\
 \noalign{\smallskip}\hline\noalign{\smallskip}
 $T$ & MJD  &  $56066.8$  &  $2.9$  \\
 $P$ & days  &  $656.0$  &  $2.3$  \\
 $a$ & mas  &  $3.44$  &  $0.18$  \\
 $e$ &   &  $0.663$  &  $0.013$  \\
 $\Omega$ & deg  &  $92.3$  &  $6.4$  \\
 $\omega$ & deg  &  $87.7$  &  $6.0$  \\
 $i$ & deg  &  $142.2$  &  $4.6$  \\
 $f_H$ &   &  $0.27$  &  $0.02$  \\
 \noalign{\smallskip}\hline\noalign{\smallskip}
 \multicolumn{4}{c}{From apparent orbit and distance}\\
 $d$ & pc  &  $1300$  &  $200$  \\
 $M_t$ & ${M}_\odot$  &  $28$  &  $13$  \\
 \noalign{\smallskip}\hline
 \end{tabular*}
 \end{minipage}
  \caption{Best fit orbital solution to the astrometric observations of CPD-47~2963. Top: motion of the secondary around the primary. The periastron of the secondary is represented by a filled symbol and the line of nodes by a dashed line. Bottom-left: best fit parameters with random noise on the dataset. Bottom-right: best fit parameters and uncertainties.}
\end{figure*}

\clearpage
\onecolumn
\section{Observation log}
\begin{longtab}\centering
\begin{longtable}{lccccccc}
\caption{\label{tab:obslog} Observation log. The separation and position angle are the position of the secondary (faintest in H band) with respect to the primary (brightest in H band), measured east (90~deg) from north (0~deg). The Night column corresponds to the UT date at Paranal at the beginning of the night (ESO archive convention). The columns emax and emin are the FWHM of the major and minor axes of the astrometric error ellipse. The column P.A.~emax is the position angle of the major axis. This table is available in electronic form at the CDS via\newline http://cdsweb.u-strasbg.fr/cgi-bin/qcat?J/A+A/}\\ \hline\hline
Target & Night & MJD & Sep. & P.A. & emax & emin & P.A.~emax \\
& & & (mas) & (deg) & (mas) & (mas) & (deg) \\ \hline
\endfirsthead
\caption{continued.}\\
\hline\hline
Target & Night & MJD & Sep. & P.A. & emax & emin & P.A.~emax \\
& & & (mas) & (deg) & (mas) & (mas) & (deg) \\ \hline
\endhead
\hline
\endfoot
 HD54662 & 2009-11-03 & 55139.341 & 3.650 & 257.800 & 0.650 & 0.150 & 133 \\
HD54662 & 2010-03-25 & 55281.055 & 3.640 & 327.360 & 0.630 & 0.460 & 157 \\
... & ... & ... & ... & ... & ... & ... & ... \\
... & ... & ... & ... & ... & ... & ... & ... \\
CPD-472963 & 2015-12-16 & 57373.211 & 1.030 & 30.760 & 0.580 & 0.360 & 138 \\
CPD-472963 & 2015-12-30 & 57387.250 & 1.030 & 329.250 & 0.450 & 0.280 & 9 \\
\noalign{\smallskip}\hline
\end{longtable}
\end{longtab}

\end{document}